\documentclass[11pt]{article}

\usepackage{jheppub}
\usepackage{amsmath,latexsym,enumerate,array}
\usepackage{slashed, mathrsfs, makeidx, hyperref}
\usepackage{graphicx,wasysym,color}
\usepackage{shuffle}

\usepackage[english]{babel}
\usepackage{url}
\usepackage{setspace}

\usepackage{float}
\pdfoutput=0 
\allowdisplaybreaks[1]

\usepackage{color}

\restylefloat{figure}
\definecolor{dgreen}{rgb}{0,0.70,0.30}
\definecolor{gold}{rgb}{0.85,.66,0}
\definecolor{purple}{rgb}{1.0,0.3,0.6}

\def\ba{\begin{array}}
\def\ea{\end{array}}

\newcommand{\bea}{\begin{eqnarray}}
\newcommand{\eea}{\end{eqnarray}}

\newcommand{\nn}{{\nonumber}}

\def\beq{\begin{equation}}
\def\eeq{\end{equation}}

\def\eqn#1{eq.~(\ref{#1})}

\newcommand{\dd}{\mathrm{d}}
\newcommand{\te}{\textrm}
\newcommand{\al}{\alpha}

\newcommand{\la}{\lambda}
\newcommand{\vph}{\varphi}
\newcommand{\ap}{{\alpha'}}

\title{Heterotic and bosonic string amplitudes via field theory}
\author[a]{Thales Azevedo,}
\author[a]{Marco Chiodaroli,}
\author[a,b]{Henrik Johansson,}
\author[c,d]{Oliver Schlotterer}

\affiliation[a]{Department of Physics and Astronomy, \\ Uppsala University, 75108 Uppsala, Sweden}
\affiliation[b]{Nordita, Stockholm University and KTH Royal Institute of Technology,\\  Roslagstullsbacken 23,
10691 Stockholm, Sweden}
\affiliation[c]{Max--Planck--Institut f\"ur Gravitationsphysik,
Albert--Einstein--Institut,
14476 Potsdam, Germany}
\affiliation[d]{Perimeter Institute for Theoretical Physics,
Waterloo, ON N2L 2Y5, Canada}
\emailAdd{thales.azevedo@physics.uu.se}
\emailAdd{marco.chiodaroli@physics.uu.se}
\emailAdd{henrik.johansson@physics.uu.se}
\emailAdd{olivers@aei.mpg.de}

\date{\today}

\abstract{Previous work has shown that massless tree amplitudes of the type I and IIA/B superstrings can be dramatically simplified by expressing them as double copies between field-theory amplitudes and scalar disk/sphere integrals, the latter containing all the $\alpha'$-corrections. In this work, we pinpoint similar double-copy constructions for the heterotic and bosonic string theories using an $\alpha'$-dependent field theory and the same disk/sphere integrals. Surprisingly, this field theory, built out of dimension-six operators such as $(D_\mu F^{\mu \nu})^2$, has previously appeared in the double-copy construction of conformal supergravity. 
We elaborate on the $\alpha' \rightarrow \infty$ limit in this picture and derive new amplitude relations for various gauge--gravity theories from those of the heterotic string.   
}

\preprint{UUITP-07/18 \\
\phantom{~} \hfill NORDITA-2018-020 \\
}

\begin{document}

\maketitle{}

\setcounter{tocdepth}{2}

\numberwithin{equation}{section}

\newpage


\section{Introduction}

Over the recent years, a surprising network of double-copy connections between seemingly-unrelated field and string theories has emerged. The most accurately studied example involves the description of gravitational scattering amplitudes as ``squares'' of suitably-arranged gauge-theory building blocks. At tree level, such  relations between gravity- and gauge-theory amplitudes have been first derived from string theory by Kawai, Lewellen and Tye (KLT) in 1986 \cite{Kawai:1985xq}. More than 20 years later, Bern, Carrasco, and one of the present authors (BCJ) obtained a more flexible formulation~\cite{BCJ} of the tree-level KLT formula that admits loop-level understanding of the gravitational double copy~\cite{loopBCJ} and links it to a duality between color and kinematics in gauge theory. The BCJ double copy has since become the state-of-the-art method for computing multiloop amplitudes in supergravity \cite{Carrasco:2011mn,Bern:2012uf,Bern:2012cd, Bern:2013uka, Bern:2014sna,Bern:2017yxu,Johansson:2017bfl, Bern:2017ucb}.

A variety of further field theories have been recently fit into double-copy patterns, including Einstein--Yang--Mills (EYM) theories \cite{Chiodaroli:2014xia, Chiodaroli:2015rdg, Chiodaroli:2017ngp}, Born--Infeld, non-linear sigma models of Goldstone bosons and special theories of Galileons \cite{Cachazo:2014xea}, large/infinite families of supergravities with reduced supersymmetry \cite{Carrasco:2012ca,Johansson:2014zca,Chiodaroli:2015wal}, and some simple examples of gauged supergravities \cite{Chiodaroli:2017ehv}. Particularly relevant to this work, the double copy has been extended to conformal supergravities \cite{Johansson:2017srf} with a construction involving a specific higher-derivative $(DF)^2$ gauge theory. The latter is built out of dimension-six operators and uniquely determined by the requirement that its amplitudes obey the color-kinematics duality.

The Cachazo--He--Yuan (CHY) formulation \cite{Cachazo:2013gna, Cachazo:2013hca, Cachazo:2013iea} of field-theory amplitudes, together with its ambitwistor-string underpinnings \cite{Mason:2013sva, Casali:2015vta}, has been a driving force to manifest or generalize double copies. This is exemplified by ref.\ \cite{Cachazo:2014nsa} for EYM and ref.\ \cite{Cachazo:2016njl} for couplings between non-linear sigma models and bi-adjoint scalars. The CHY formalism has been used to derive amplitude relations between EYM and gauge theories \cite{Nandan:2016pya, delaCruz:2016gnm, Teng:2017tbo, Du:2017gnh}, new representations of loop amplitudes in super--Yang--Mills (SYM) and supergravity \cite{Geyer:2015bja, Geyer:2015jch, Geyer:2016wjx} as well as a one-loop incarnation of the KLT formula \cite{He:2016mzd, He:2017spx}.

String theories with a finite tension $\ap^{-1}$ are a particularly well-established framework to obtain a double-copy perspective on field-theory amplitudes. Apart from the original tree-level KLT formula \cite{Kawai:1985xq}, the BCJ relations \cite{BCJ} between color-ordered gauge-theory trees were elegantly proven through the field-theory limit $\ap \rightarrow 0$ of the monodromy relations \cite{BjerrumBohr:2009rd, Stieberger:2009hq} between open-string amplitudes (see \cite{Tourkine:2016bak, Hohenegger:2017kqy, Ochirov:2017jby} for recent extensions to loop level). Also, EYM amplitude relations naturally descend from type-I superstrings \cite{Stieberger:2016lng} and heterotic strings \cite{Schlotterer:2016cxa}. Furthermore, string-theory methods allowed for the first systematic construction of gauge-theory amplitude representations where the color-kinematics duality is manifest: So far, this has been achieved at tree level \cite{Mafra:2011kj}, one loop \cite{Mafra:2014gja, He:2015wgf} and two loops \cite{Mafra:2015mja} with maximal supersymmetry. Also, see refs.~\cite{Berg:2016wux, Berg:2016fui} for one-loop advances at reduced supersymmetry and \cite{He:2017spx} for cross-fertilization with the CHY formalism.

A remarkable result is that even the {\it open} superstring can be identified as a double copy once the polarization dependence of its tree-level amplitudes is expressed via gauge-theory trees \cite{Mafra:2011nv}. In a suitable organization of the accompanying $\ap$-dependent worldsheet integrals, this decomposition of open-string amplitudes follows the pattern of the field-theory KLT formula \cite{Zfunctions}. This unexpected double-copy structure of the open superstring suggests an interpretation of the moduli-space integrals over disk worldsheets as scattering amplitudes in a putative effective theory of scalars. The latter has been dubbed Z-theory and investigated in view of its low-energy limit \cite{Carrasco:2016ldy, Carrasco:2016ygv} where couplings between non-linear sigma models and bi-adjoint scalars are recovered as well as the $\ap$-expansion of its non-linear field equations \cite{Mafra:2016mcc}. The Z-theory approach \cite{Carrasco:2016ldy, Carrasco:2016ygv} underpins the color-kinematics duality of non-linear sigma models \cite{Chen:2013fya, Cachazo:2014xea} from a string-theory perspective and gives rise to the same amplitude representations as the cubic action of Cheung and Shen \cite{Cheung:2016prv}. According to the evidence presented in ref.~\cite{Mafra:2017ioj}, the double-copy structure of the open superstring is expected to extend to loop level.

Tree-level amplitudes of the open bosonic string are conjectured to admit the same double-copy organization as observed for the superstring \cite{Huang:2016tag}. In a universal basis of disk integrals or Z-theory amplitudes, the only difference between bosonic and supersymmetric strings resides in the rational kinematic factors; these are $\ap$-independent gauge-theory trees in the case of the superstring \cite{Mafra:2011nv}, and more complicated rational functions of $\ap$ in the case of the bosonic string \cite{Huang:2016tag}. As a main result of this work, these $\ap$-dependent kinematic factors of the bosonic string will be reproduced from a specific field-theory Lagrangian, thus pinpointing the missing double-copy component of the open bosonic string. Surprisingly, this field theory, dubbed $(DF)^2\,{+}\, \te{YM}$, turns out to be a massive deformation of the $(DF)^2$ theory entering the double-copy construction for conformal supergravity \cite{Johansson:2017srf}. More precisely, the $(DF)^2\, {+}\, \te{YM}$ theory, depending on $\ap$ through its mass parameter, interpolates between pure Yang--Mills (YM) theory ($\ap \rightarrow 0$) and the undeformed $(DF)^2$ theory ($\ap \rightarrow \infty$).
\begin{table}[h]
\beq \! \! \!
{\setstretch{1.75} 
\begin{array}{c||c|c|c}
{\rm string}\otimes {\rm QFT}  &\te{SYM} &(DF)^2 \,{+}\, \te{YM}&\ (DF)^2\, {+}\, \te{YM}\,{+}\,\phi^3 \\\hline \hline
\te{Z-theory} \ & \ \te{open superstring} \ \, & \, \te{open bosonic string} \, &{\setstretch{0.85} \begin{array}{c} 
\te{compactified open} \\
  \te{bosonic string}
  \end{array}}  \\
\te{sv}(\te{open superstring})  \ & \ \te{closed superstring}  \ \, &\ \te{heterotic (gravity)} \ \, &\, \te{heterotic\,(gauge/gravity)}  \\
\te{sv}(\te{open\,bosonic\,string})  \, &  \, \te{heterotic\,(gravity)} \, & \, \te{closed\,bosonic\,string} \, &{\setstretch{0.85} \begin{array}{c} 
\te{compactified closed} \\
  \te{bosonic string}
  \end{array}}
\end{array}} \nonumber
\eeq
\caption{Various known double-copy constructions of string amplitudes. The last two columns of the table are completed in this work. The single-valued projection sv($\bullet$) converts disk to sphere integrals. }
\label{overview}
\end{table}

In addition to the double-copy representations for open-string amplitudes (first row of table \ref{overview}), we will identify similar field-theory constituents for gravitational amplitudes of the closed bosonic and the heterotic string (second and third row, table \ref{overview}). This will generalize a simplified representation of massless tree amplitudes of the closed superstring where the cancellations of various multiple zeta values in the $\ap$-expansion are taken into account: The selection rules applied to the multiple zeta values in passing from the open to the closed superstring \cite{Schlotterer:2012ny} are mathematically formalized by the so-called single-valued ``sv''  projection \cite{Schnetz:2013hqa, Brown:2013gia}. The closed superstring emerges from the double copy between SYM and the sv-projected open superstring \cite{Stieberger:2013wea}. As pointed out in ref.~\cite{Huang:2016tag}, gravitational amplitudes of superstrings, heterotic strings and bosonic strings only differ in their field-theory kinematic factors. Hence, different double copies involving one of SYM or $(DF)^2\, {+}\, \te{YM}$ as well as a single-valued open-string theory gives rise to the closed bosonic string or the gravity sector of the heterotic string, see table \ref{overview}.

As a second major novel result, we extend the collection of field-theory double copies to tree-level amplitudes involving any number of gauge- and gravity multiplets in the heterotic string. For this purpose, bi-adjoint scalars are incorporated into the $\ap$-dependent $(DF)^2\, {+}\, \te{YM}$ theory, resulting in the so-called $(DF)^2 {+} \te{YM}{+}\phi^3$ theory whose Lagrangian will be given below. Upon double copying with the single-valued open superstring, this completes the massless sector of the heterotic string, see the last column of table \ref{overview}.
The remaining double-copy options involving $(DF)^2 {+} \te{YM}{+}\phi^3$ theory and one of Z-theory or the single-valued open bosonic string yield compactified versions of bosonic string theories\footnote{The compactification geometries can be taken to be two independent copies of the same sixteen-dimensional internal manifolds as used for the non-supersymmetric chiral half of the heterotic string \cite{Gross:1985fr, Gross:1985rr}, giving the Lie group $SO(32)$ or $E_8 \times E_8$. More generally, the color degrees of freedom can be generalized to arbitrary Lie groups since the tree-level amplitudes under discussion are not yet constrained by modular invariance.}.

In the low-energy limit $\ap \rightarrow 0$, the YM$+\phi^3$ theory, known from the double copy of EYM \cite{Chiodaroli:2014xia, Chiodaroli:2015rdg}, can be obtained from the $(DF)^2 {+} \te{YM}{+}\phi^3$ theory. 
Hence, our double-copy construction of the heterotic string encodes all EYM amplitude relations that reduce multi-trace couplings of gluons and gravitons to pure gauge-theory amplitudes and should reproduce the results of ref.~\cite{Du:2017gnh}.  The opposite limit $\ap\rightarrow \infty$ of the $(DF)^2 {+} \te{YM}{+}\phi^3$ theory, on the other hand, gives access to conformal supergravity coupled to gauge bosons upon double copy with SYM. Indeed, in this regime, the double-copy amplitudes built from the $(DF)^2 {+} \te{YM}{+}\phi^3$ theory and SYM agree with those coming from the heterotic ambitwistor string, which were studied in detail in ref.\ \cite{Azevedo:2017lkz}. 

Finally, we will deduce from the CHY formulae, given in that reference, that amplitude relations of conformal supergravity coupled to SYM are recovered in the $\ap \rightarrow \infty$ limit of the corresponding amplitude relations of the (conventional) heterotic string \cite{Schlotterer:2016cxa}. An alternative way of arriving at conformal-gravity amplitudes via an $\ap\to\infty$ limit involves the twisted-string construction of Huang, Siegel and Yuan\ \cite{Huang:2016bdd}. We show that taking this limit for the twisted heterotic-string amplitudes one obtains precisely a KLT formula with the undeformed $(DF)^2$ theory and SYM amplitudes.

This paper is organized as follows: A review in section \ref{sec2} sets the stage for introducing the $(DF)^2\, {+}\, \te{YM}$ theory and for explaining its significance for the open and closed bosonic string in section \ref{sec3}. The generalization of these results to the heterotic string is discussed in section \ref{sec4} which allows to explore various corollaries for different types of gauge-gravity theories in the $\ap\rightarrow \infty$ limit in section \ref{sec5}.

\section{Review}
\label{sec2}

In this section, we review the known double-copy structure of tree-level amplitudes in various string theories and conformal 
supergravity. This will set the stage for sections~\ref{sec3} and \ref{sec4} where we identify the field theory that governs the polarization dependent kinematic factors of the bosonic and heterotic string amplitudes. For simplicity, throughout this paper the field- and string-theory amplitudes will be given without overall coupling-dependent prefactors. Appropriate powers of the gauge coupling $g$ and the gravitational coupling $\kappa$ can be restored following appendix \ref{appNorm}.

\subsection{Open superstrings}
\label{sec21}

The manifestly supersymmetric pure-spinor formalism \cite{Berkovits:2000fe} has been used to determine
all tree amplitudes of massless open-superstring states \cite{Mafra:2011nv} in terms of simpler building blocks,
\beq
 {\cal A}_{\rm s}(1,\pi(2,3,\ldots,n{-}2),n{-}1,n) = \sum_{\rho \in S_{n-3}} F_\pi{}^\rho A_{\rm SYM}(1,\rho(2,3,\ldots,n{-}2),n{-}1,n) \, .
 \label{2.1}
\eeq
The sum runs over $(n{-}3)!$ permutations $\rho$, and the factor of $A_{\rm SYM}$ refers to color-ordered tree amplitudes of ten-dimensional SYM \cite{Mafra:2010jq} which carry all of the polarization dependence and supersymmetry. The $\ap$-dependence, by contrast, is entirely carried by the integrals \cite{Mafra:2011nv} 
\beq
F_\pi{}^\rho =(2 \ap)^{n-3} \! \! \! \! \! \! \! \! \! \! \! \! \! \! \! \! \! \! \! \! \int \limits_{0 \leq z_{2_\pi}\leq z_{3_\pi} \leq \ldots \leq z_{(n{-}2)_\pi}\leq 1} \! \! \! \! \! \! \! \! \! \! \! \! \! \! \! \! \! \! \! \! \dd z_2 \, \dd z_3 \, \ldots\, \dd z_{n-2} \ \prod_{i<j}^{n-1} |z_{ij}|^{2\alpha' s_{ij}} \rho \, \bigg\{
\prod_{k=2}^{n-2} \sum_{m=1}^{k-1} \frac{ s_{mk}}{z_{km}}
\bigg\} 
\label{2.2}
\eeq
over moduli spaces of punctured disks which are taken to be in the ${\rm SL}(2,\mathbb R)$ gauge with $(z_1,z_{n-1},z_n)=(0,1,\infty)$. As is well known, the $A_{\rm SYM}$ obeys BCJ relations \cite{BCJ} and $ {\cal A}_{\rm s}$ obeys monodromy relations \cite{BjerrumBohr:2009rd, Stieberger:2009hq}. The permutations $\rho,\pi \in S_{n-3}$ are chosen such that 
the disk integrals in (\ref{2.2}) form independent $(n{-}3)!$-element bases under the BCJ and monodromy relations, respectively. Thus, the $F_\pi{}^\rho $ form an $(n{-}3)! \times (n{-}3)!$ matrix with entries indexed by integration domains $\pi$ and permutations $\rho$ of integrand labels. The permutations $\rho$ are understood to act on the insertion labels as\footnote{Note that we use the usual string-theory convention where $s_{ij}$ are defined to be half of the value used for field-theory Mandelstam variables.} 
\beq
\rho\big\{s_{ij} \big\} \equiv k_{\rho(i)} \cdot k_{\rho(j)}  \ , \ \ \ \ \ \ \rho\big\{z_{ij} \big\}  \equiv z_{\rho(i)} - z_{\rho(j)} \ ,
\label{2.2z}
\eeq
where $\rho$ acts trivially on the labels $1, n{-}1, n$ of the fixed punctures.

Alternatively, one can organize the disk integrals in terms of the so-called Z-theory amplitudes\footnote{The inverse factor of ${\rm vol}({\rm SL}(2,\mathbb R))$ refers to the standard procedure of fixing three punctures $z_i,z_j,z_k\rightarrow (0,1,\infty) $ while trading the integration measure $\dd z_i \, \dd z_j \, \dd z_k$ for a Jacobian $|z_{ij}z_{ik} z_{jk}|$. Accordingly, the three-point instances of the disk integrals (\ref{2.2b}) trivialize to $Z(1,2,3|1,2,3)=1 $ and $Z(1,2,3|1,3,2)=-1$.},
\beq
Z(\pi(1,2,\ldots,n) \, |\, \rho(1,2,\ldots,n)) = (2 \ap)^{n-3} \! \! \! \! \! \! \! \! \! \! \! \! \! \! \! \! \! \! \! \!  \! \! \! \!\int \limits_{\pi \, \{  - \infty \leq z_{1}\leq z_{2} \leq \ldots \leq z_{n}\leq \infty \}} \! \! \! \! \! \! \! \! \! \! \! \! \! \! \! \! \! \! \! \! \frac{\dd z_1 \, \dd z_2 \, \ldots\, \dd z_{n}}{{\rm vol}({\rm SL}(2,\mathbb{R}))} \ 
\frac{ \prod_{i<j}^{n} |z_{ij}|^{2\alpha' s_{ij}}  }{ \rho \, \{ z_{12} z_{23} \cdots z_{n-1,n} z_{n,1} \} } \, .
\label{2.2b}
\eeq
The massless open-string amplitudes (\ref{2.1}) can then be written in a more suggestive form \cite{Broedel:2013aza}
\begin{align}
 {\cal A}_{\rm s}(\pi(1,2,3,\ldots,n)) &= \sum_{\tau,\rho}  Z(\pi(1,2,\ldots,n) \, | \, 1,\tau(2,3,\ldots,n{-}2),n,n{-}1)  \notag\\
 & \ \ \ \ \times S[\tau | \rho]_1  A_{\rm SYM}(1,\rho(2,3,\ldots,n{-}2),n{-}1,n) \,,
 \label{2.2c}
\end{align}
where the summation range $\tau,\rho \in S_{n-3}$ is suppressed here and in later equations for ease of notation. 
This form is isomorphic to the field-theory KLT formula, and the symmetric $(n-3)!$-by-$(n-3)!$ matrix  $S[\tau|\rho]_1 = S[\tau(2,\ldots,n{-}2) |\rho (2,\ldots,n{-}2)]_1$ is the field-theory KLT kernel. For a fixed choice of basis, the KLT kernel was first written down to all multiplicity in ref.~\cite{Bern:1998sv} and later extended to more general basis choices in refs.~\cite{BjerrumBohr:2010ta, BjerrumBohr:2010yc, momentumKernel}. Here we give a compact recursive definition\footnote{Note that the KLT kernel used here is $-i(-2)^{3-n}$ times the one defined in ref.~\cite{Bern:1998sv}.}~\cite{Carrasco:2016ldy}, 
\beq
S[A,j|B,j,C]_1 = (k_1+k_B)\cdot k_j  \, S[A|B,C]_1  \ , \ \ \ \ \ \ S[2|2]_1 = s_{12} \ ,
\label{2.6}
\eeq
where multiparticle labels such as $B=(b_1,b_2,\ldots,b_p)$ encompass multiple external legs and $k_B = k_{b_1}+k_{b_2}+\ldots+k_{b_p}$ denotes their region momentum. 

Since the open-string amplitudes in eqs.~(\ref{2.1}) and (\ref{2.2c}) are expressed in a basis of SYM amplitudes, the corresponding coefficients are unique and thus equal to each other \cite{Broedel:2013aza},
\beq
 F_\pi{}^\rho = \sum_{\tau } Z(1,\pi(2,3,\ldots,n{-}2),n{-}1,n \, | \, 1,\tau(2,3,\ldots,n{-}2),n,n{-}1) S[\rho | \tau]_1  \, .
\label{2.15}
\eeq
For theories where all-multiplicity tree-level amplitudes can be expressed using the KLT formula, we say that the theory (or a sector of a theory) is constructible as a {\it double copy} out of a left and right theory.  It is convenient to use a word formula, such as 
\beq
(\text{massless open superstring}) = (\text{Z-theory}) \otimes  ({\rm SYM})\,,
\label{OS_DC}
\eeq
that concisely expresses this statement. This context-free formula is useful since the double copy can be expressed in many 
different ways (for example, using the BCJ or CHY frameworks), the KLT formula being just one instance of this. The slightly
abusive specifier ``massless open superstring'' means that we only allow for massless external modes, although
massive open-string vibration modes still contribute to (\ref{2.2}) and (\ref{2.2b}) via internal propagation. Both (\ref{OS_DC}) and later double-copy statements on string amplitudes are understood to only apply to massless external states, and we will suppress the specifier ``massless'' henceforth.

Whether the Z-theory amplitudes (\ref{OS_DC}) define a sensible theory or not is an interesting open question. Evidence supporting an interpretation as a theory stems from the observation that, at leading order in $\ap$, the disk integrals (\ref{2.2b}) reproduce the doubly-partial amplitudes of non-linear sigma models coupled to bi-adjoint scalars upon (partial) symmetrization in $\pi$ 
\cite{Carrasco:2016ldy, Carrasco:2016ygv}. Hence, the $Z(\pi|\rho)$ integrals are interpreted as the doubly-partial 
amplitudes in a collection of effective scalar theories dubbed Z-theory which captures all the $\ap$-dependence of 
open-string amplitudes. The low-energy expansion of the Z-theory equations of motion can be found in ref.~\cite{Mafra:2016mcc}.

\subsection{Closed superstrings}
\label{sec21closed}

For massless (gravitational) tree amplitudes of the closed type-IIA/B superstring, combining two copies of the open-string worldsheet integrands in (\ref{2.1}) leads to the structure \cite{Schlotterer:2012ny},
\begin{align}
{\cal M}_{\rm II} &= \sum_{\tau,\rho,\sigma} \tilde A_{\rm SYM}(1,\tau,n,n{-}1) S[\tau|\rho]_1 
({\rm sv} F)_{\rho}{}^\sigma A_{\rm SYM}(1,\sigma,n{-}1,n) \ ,
\label{2.5}
\end{align}
where the field-theory KLT kernel is given by (\ref{2.6}) and the type-IIA/B amplitudes are obtained from each other by flipping the relative chirality of the ten-dimensional ${\cal N}=(1,0)$ SYM theories.
The summation ranges for permutations $\tau,\rho,\sigma \in S_{n-3}$ of $2,3,\ldots,n{-}2$ are again
suppressed in (\ref{2.5}) and later equations.

The notation ${\rm sv} F$ in (\ref{2.5}) refers to the single-valued projection \cite{Schnetz:2013hqa, Brown:2013gia} of multiple zeta values (MZVs)
\beq
\zeta_{n_1,n_2,\ldots,n_r} \equiv \sum_{0<k_1<k_2<\ldots<k_r} k_1^{-n_1} k_2^{-n_2} \ldots k_r^{-n_r} \ , \ \ \ \ \ \ n_r \geq 2 \ ,
\label{2.3}
\eeq
which arise in the low-energy expansion of open- and closed-string tree amplitudes \cite{Aomoto, Terasoma, Brown:2009qja, Stieberger:2009rr, Schlotterer:2012ny}. More precisely, the $w^{\rm th}$ order in the $\ap$-expansion of the functions $F_\pi{}^\rho$ in (\ref{2.2}) exclusively involves MZVs (\ref{2.3}) of weight $w=n_1+n_2+\ldots+n_r$, so the $F_\pi{}^\rho$ are said to exhibit uniform transcendentality. The matrix $({\rm sv} F)_{\rho}{}^\sigma$ in (\ref{2.5}) is built from moduli-space integrals over punctured spheres whose $\ap$-expansion can be conjecturally obtained from the open-string functions $F_{\rho}{}^\sigma$ in (\ref{2.2}) through the formal map \cite{Schlotterer:2012ny, Stieberger:2013wea}
\beq
 {\rm sv}(\zeta_2)=0 \ , \ \ \ \ \ \ 
 {\rm sv}(\zeta_{2k+1})=2\zeta_{2k+1} \ , \ \ \ \ \ \ 
 {\rm sv}(\zeta_{3,5})=-10\zeta_3\zeta_5 \ , \ \ \ \ \ \ \ldots
\label{svmap}
\eeq
acting on the MZVs in their low-energy expansion. The terminology for the single-valued projection ``sv'' is motivated  by the realization of ${\rm sv}(\zeta_{n_1,n_2,\ldots,n_r})$ via single-valued polylogarithms at unit argument \cite{svpolylog, Schnetz:2013hqa, Brown:2013gia}, see the references for a generalization of (\ref{svmap}) to higher weight and depth. The single-valued projection (\ref{svmap}) preserves uniform transcendentality and the product structure, $ {\rm sv}(\zeta_{m_1,m_2,\ldots}\zeta_{n_1,n_2,\ldots})= {\rm sv}(\zeta_{m_1,m_2,\ldots}) {\rm sv}(\zeta_{n_1,n_2,\ldots})$. Further details on the interplay of $\ap$-expansions with the single-valued map are reviewed in appendix \ref{appD}. First evidence for a similar correspondence between one-loop amplitudes of open and closed strings has been given in \cite{Broedel:2018izr}.

Given that the factors $A_{\rm SYM}$ in the open-string amplitude (\ref{2.1}) are unaffected by the ${\rm sv}$ operation, the closed-string amplitude (\ref{2.5}) can be cast into a double-copy form~\cite{Stieberger:2013wea}
\begin{align}
{\cal M}_{\rm II} &= \sum_{\tau,\rho}   \tilde A_{\rm SYM}(1,\tau ,n,n{-}1) S[\tau|\rho]_1 {\rm sv}{\cal A}_{\rm s}(1,\rho,n{-}1,n)  \ ,
\label{svformula}
\end{align}
where ${\rm sv}$ only acts on the open-string amplitudes.  The double copy for the closed-string amplitude is then summarized by the word equation
\beq
(\text{closed superstring})= ({\rm SYM}) \otimes  {\rm sv}\big(\textrm{open superstring} \big) \, ,
\label{CSvsOS}
\eeq
which is understood to apply to massless external states according to our earlier disclaimer. Alternatively, the double copy 
can be written more symmetrically by substituting \eqn{OS_DC} into \eqn{CSvsOS}, giving a ``triple copy'' 
\beq
(\text{closed superstring})= ({\rm SYM}) \otimes  {\rm sv}\text{(Z-theory)}  \otimes ({\rm SYM})\,.
\label{CSvsOSprime}
\eeq
The corresponding KLT-like formula is
\begin{align}
{\cal M}_{\rm II} &= \! \! \sum_{\tau,\pi, \sigma, \rho} \! \!  \tilde A_{\rm SYM}(1,\tau ,n,n{-}1)  S[\tau|\pi]_1 {\rm sv} Z(1,\pi,n{-}1,n | 
1,\sigma,n,n{-}1) S[\sigma|\rho]_1 A_{\rm SYM}(1,\rho,n{-}1,n) \, .
\label{kltlike}
\end{align}
However, note that this ``triple copy'' is not a new concept, it simply amounts to applying the double copy twice. Note that ${\rm sv} Z(\pi|\sigma)$ can be interpreted as closed-string versions of Z-theory amplitudes \cite{Mafra:2016mcc}, and they are given by integrals \cite{Stieberger:2014hba}
\beq
{\rm sv} \, Z(\tau  | \sigma) =  \left( \frac{2\ap}{\pi } \right)^{n-3}  \! \! \!  \int  \frac{\dd^2 z_1 \, \dd^2 z_2 \, \ldots\, \dd^2 z_{n}}{{\rm vol}({\rm SL}(2,\mathbb{C}))} \ 
\frac{ \prod_{i<j}^{n} |z_{ij}|^{4\alpha' s_{ij}}  }{ \tau \, \{ \bar{z}_{12} \bar{z}_{23} \cdots \bar{z}_{n-1,n} \bar{z}_{n,1} \}  \sigma \, \{ z_{12} z_{23} \cdots z_{n-1,n} z_{n,1} \} } \,,
\label{2.2bcl}
\eeq
over the moduli space of punctured Riemann spheres\footnote{For the complex punctures $z_i$ of (\ref{2.2bcl}),
the inverse factor of ${\rm vol}({\rm SL}(2,\mathbb C))$ amounts to fixing three punctures $z_i,z_j,z_k\rightarrow (0,1,\infty) $ 
while trading the integration measure $\dd^2 z_i \, \dd^2 z_j \, \dd^2 z_k$ for a Jacobian $|z_{ij}z_{ik} z_{jk}|^2$. The exact closed-string normalization of $\alpha'$ can be attained by replacing $\alpha' \rightarrow \frac{\ap}{4}$ in (\ref{2.2bcl}) and later equations on type-II, heterotic-string and closed-bosonic-string amplitudes. We do not track this rescaling of $\alpha'$ between open- and closed-string setups to avoid the ubiquity of denominators.}.

\subsection{Bosonic and heterotic strings}
\label{sec22}

For the open bosonic string, $n$-point tree amplitudes were conjectured\footnote{The expectation for (\ref{2.8}) to hold at all multiplicities $N$ is based on the explicit checks at $N\leq 7$ \cite{Huang:2016tag}.} to have the same expansion in terms of a basis of disk integrals (\ref{2.2}) as for the
 open superstring \cite{Huang:2016tag},
\beq
 {\cal A}_{\rm bos}(1,\pi,n{-}1,n) = \sum_{\rho}  F_\pi{}^\rho \, B(1,\rho,n{-}1,n) \, .
 \label{2.8}
\eeq
The kinematic factors $B$, which will soon be identified with partial amplitudes of a specific gauge theory, capture the entire dependence on the external polarizations, and they additionally depend on $\ap$. We give the lowest-multiplicity expressions in terms of gluon polarizations $e^\mu_i$ and linearized field strengths $f^{\mu \nu}_i = k^\mu_ie^\nu_i-e^\mu_ik^\nu_i$ with $D$-dimensional Lorentz indices ($D=26$ for the critical string) and $i=1,2,\ldots,n$ labelling the external legs. We use shorthands $f_{ij} \equiv \frac{1}{2} f_i^{\mu \nu} f_j^{\mu \nu}= s_{ij}(e_i{\cdot} e_j) - (k_i{\cdot} e_j)(k_j{\cdot} e_i)$ and $g_i \equiv (k_{i-1} {\cdot} e_i) s_{i,i+1} - (k_{i+1}{\cdot} e_i) s_{i-1,i}$ in
\begin{align}
B(1,2,3) &= A_{\rm YM}(1,2,3) +4 \alpha' (e_1 \cdot k_2)(e_2 \cdot k_3)(e_3\cdot k_1) \label{2.9}
\\
 \! \! \! B(1,2,3,4) &= A_{\rm YM}(1,2,3,4) + 4\alpha' s_{13} \Big\{ \Big[ \frac{ f_{12} f_{34}  }{s_{12}^2 (1{-}2\alpha' s_{12})} {+} {\rm cyc}(2,3,4) \Big] - \frac{ g_1 g_2 g_3 g_4 }{s_{12}^2 s_{13}^2 s_{23}^2} 
\Big\}  \ ,\label{2.10}
\end{align}
and the YM tree-level amplitudes are normalized as exemplified by the three-point case, 
\beq
A_{\rm YM}(1,2,3) = 2
(e_1\cdot e_2)(k_1\cdot e_3) + {\rm cyc}(1,2,3) \ .
 \label{YM3pt}
\eeq
While $B(1,2,3,4,5)$ at five points can be obtained from ref.~\cite{Huang:2016tag}, extracting the higher-multiplicity generalizations\footnote{The subleading order of $B(1,2,\ldots,n)$ in $\ap$ is governed by a deformation of SYM through a ${\rm Tr}(F^3)$-operator whose $D$-dimensional $n$-point tree amplitudes have been considered in ref.~\cite{He:2016iqi}.} from string theory is an open problem. Instead, we will later on employ field-theory techniques to obtain an all-multiplicity description of the kinematic factors $B$ in the bosonic-string amplitude (\ref{2.8}).

Given that the geometric-series expansion of the factor $(1-2\ap s_{ij})^{-1}$ in (\ref{2.10}) leads to purely rational coefficients for $(\ap s_{ij})^w$, the amplitudes (\ref{2.8}) of the open bosonic string violate uniform transcendentality. Instead, the combination of the purely rational $B$ and the uniformly transcendental integrals $F_\pi{}^\rho$ in (\ref{2.2}) sets the upper bound that the order of $\ap^w$ involves MZV of weight $\leq w$. In particular, given the leading low-energy behavior
\beq
B(1,2,\ldots,n) = A_{{\rm YM}}(1,2,\ldots,n) + {\cal O}(\ap)
\label{2.11}
\eeq
imposed by the field-theory limit of (\ref{2.8}), the pieces of maximal transcendentality in bosonic open-string amplitudes, i.e.\ the contributions of weight-$w$ MZVs at the order of $\ap^w$, coincide with those of the superstring (\ref{2.1}) \cite{Huang:2016tag}.

Of course, the worldsheet integrand of the open bosonic string underlying (\ref{2.8}) can be exported to the gravitational sector of the heterotic string. Hence, the $n$-point tree amplitude of supergravity states $h$ in the heterotic theory \cite{Huang:2016tag} is given by the double copy
\begin{align}
{\cal M}^{h}_{\rm het} &= \sum_{\tau,\rho} B(1,\tau,n,n{-}1) S[\tau|\rho]_1  
{\rm sv} {\cal A}_{\rm s}(1,\rho,n{-}1,n)\,,
\label{2.12}
\end{align}
which is related to the type-II result (\ref{svformula}) by the replacement $\tilde A_{\rm SYM} \rightarrow B$, with the identical sphere integrals that yield single-valued open-string amplitudes (\ref{svmap}). A further replacement ${\rm sv} {\cal A}_{\rm s}\rightarrow {\rm sv} \tilde {\cal A}_{\rm bos}$ maps the heterotic-string amplitude (\ref{2.12}) to the $n$-point tree amplitudes of the closed bosonic string \cite{Huang:2016tag},
\begin{align}
{\cal M}_{\rm bos} &= \sum_{\tau,\rho} B(1,\tau,n,n{-}1) S[\tau|\rho]_1  
{\rm sv} \tilde {\cal A}_{\rm bos}(1,\rho,n{-}1,n) \, .
\label{2.12prime}
\end{align}
From the absence of MZVs in the $\ap$-expansion of $B$, the piece of maximal transcendentality at a given order in $\ap$ is universal to gravitational amplitudes in the closed superstring, the heterotic string and the closed bosonic string \cite{Huang:2016tag}.

The structure of the bosonic- and heterotic-string amplitudes (\ref{2.8}), (\ref{2.12}) and (\ref{2.12prime}) matches
the open- and closed-superstring results for ${\cal A}_{\rm s}$ and ${\cal M}_{\rm II}$. Hence, the results reviewed in this
section imply that the gravity sector of the bosonic and heterotic string exhibits a double-copy structure. Indeed, we will
pinpoint a gauge theory in section \ref{sec3} which furnishes the bosonic double-copy component and takes the role
of SYM in (\ref{OS_DC}) and (\ref{CSvsOS}).

\subsection{Field-theory BCJ relations for string amplitudes}
\label{sec24}

As a common property of the constituents in the above string double copies, they are all required to obey the BCJ relations of gauge-theory tree amplitudes \cite{BCJ}, e.g. 
\beq
\sum_{j=2}^{n-1} k_1 \cdot (k_2{+}k_3{+}\ldots{+}k_j) A_{\rm SYM}(2,3,\ldots,j,1,j{+}1,\ldots,n) =0 \, .
\label{2.13}
\eeq
Given that total derivatives in the worldsheet punctures integrate to zero,
the same relations hold for different choices of the right argument $\rho$ in the Z-theory amplitudes $Z(\pi|\rho)$ in (\ref{2.2b}) with a fixed integration domain $\pi$ \cite{Zfunctions}. Accordingly, the single-valued Z-theory amplitudes (\ref{2.2bcl}) obey BCJ relations for either the left or right argument.

For open-string amplitudes ${\cal A}_{\rm s}(\pi)$ or ${\cal A}_{\rm bos}(\pi)$, different disk orderings $\pi$ are constrained by the generalization of the BCJ relations (\ref{2.13}) to monodromy relations \cite{BjerrumBohr:2009rd, Stieberger:2009hq}. They augment the BCJ relations by a series expansion in $\zeta_2 (\ap s_{ij})^2$ which preserve the uniform transcendentality of the superstring. As firstly pointed out in ref.~\cite{Broedel:2012rc}, monodromy relations can be projected separately to any order in $\ap$ and to any MZV which is (conjecturally) independent over $\mathbb Q$. Truncating the monodromy relations among (\ref{2.8}) to zero transcendentality leads to the BCJ relations \cite{Huang:2016tag}
\beq
\sum_{j=2}^{n-1} k_1 \cdot (k_2{+}k_3{+}\ldots{+}k_j) \, B(2,3,\ldots,j,1,j{+}1,\ldots,n) =0 \ ,
\label{2.16}
\eeq
valid to all orders in $\ap$, as can be easily verified for the four-point example in (\ref{2.10}). The same reasoning implies that also single-valued open-string amplitudes satisfy the BCJ relations \cite{Stieberger:2014hba}
\beq
\sum_{j=2}^{n-1} k_1 \cdot (k_2{+}k_3{+}\ldots{+}k_j)  \, {\rm sv}{\cal A}_{{\rm s},  {\rm bos}}(2,3,\ldots,j,1,j{+}1,\ldots,n) =0 
\label{2.16prime}
\eeq
to all orders in $\ap$, ensuring consistency of the double-copy formulae (\ref{svformula}), (\ref{2.12}) and (\ref{2.12prime}).

\subsection{Conformal supergravity as a double copy}
\label{sec23}

Conformal-supergravity (CSG) amplitudes of the non-minimal Berkovits--Witten-type theory \cite{Berkovits:2004jj} are constructed from the double copy \cite{Johansson:2017srf}
\beq
{\rm CSG}=\big(\textrm{$(DF)^2$-theory}\big)\otimes \big({\rm SYM}\big)\,.
\label{CG_DC}
\eeq
Here, in the second factor, SYM stands for any pure super-Yang--Mills theory in dimensions $D\le10$, and the first factor is a bosonic gauge theory with the following Lagrangian in any dimension:
\beq
{\cal L}_{(DF)^2}=   \frac{1}{2}(D_{\mu} F^{a\, \mu \nu})^2 -  \frac{g}{3}   F^3  +\frac{1}{2}(D_{\mu} \varphi^{\alpha})^2 + \frac{g}{2} \,  C^{\alpha ab}  \varphi^{ \alpha}   F_{\mu \nu}^a F^{b\, \mu \nu }  + \frac{g}{3!}  \, d^{\alpha \beta \gamma}   \varphi^{ \alpha}  \varphi^{ \beta} \varphi^{ \gamma}\,.
\label{LagrDF}
\eeq
The field strength and covariant derivatives are defined as
\begin{align}
F_{\mu \nu}^a &= \partial_{\mu} A_{\nu}^a-\partial_{\nu} A_{\mu}^a + g  f^{abc} A_{\mu}^b A_{\nu}^c\,, \notag  \\
D_{\mu} \varphi^\alpha & = \partial_{\mu} \varphi^\alpha -  i g  (T_{R}^{a})^{\alpha \beta} A_{\mu}^a \varphi^\beta\,,  \label{fieldDef}  \\
D_{\rho} F_{\mu \nu}^a &= \partial_{\rho} F_{\mu \nu}^a +  g  f^{abc} A_{\rho}^b F_{\mu \nu}^c\,, \notag\\
F^3 &=  f^{abc} F_{\mu}^{a\, \nu}F_{\nu}^{b \, \la} F_{\la }^{c\, \mu}   \, .
\notag
\end{align}
The vector $A_{\mu}^a$ transforms in the adjoint representation of a gauge group $G$ with indices $a,b,c,\ldots$, the scalar $\varphi^\alpha$ transforms in a real representation $R$ of the gauge group (with generators $T_{R}^{a}$), and $g$ is the coupling constant\footnote{Note that the conformal supergravities do not inherit the explicit dependence on the gauge group and coupling, these are only introduced in order to define the non-abelian gauge theory. Only the kinematic structure of the gauge theory is transferred to the conformal supergravities through the double copy.}.
The Clebsch--Gordan coefficients $C^{\alpha ab}$ and $d^{\alpha \beta \gamma}$ are implicitly defined through the two relations
\bea
&&C^{\alpha ab}C^{\alpha cd} = f^{ace}f^{edb}+ f^{ade}f^{ecb}\,, \label{ClebshGordan}  \\
&&C^{\alpha ab}d^{\alpha \beta \gamma}= (T_{R}^a)^{\beta \alpha} (T_{R}^b)^{\alpha \gamma}+ C^{\beta ac} C^{\gamma cb} + (a \leftrightarrow b)\,. \nn
\eea
Requiring that the tensors $C^{\alpha ab}$, $d^{\alpha \beta \gamma}$ and $T_{R}^{a}$ transform covariantly under infinitesimal gauge-group rotations implies three additional Lie-algebra relations. Together, these five tensor relations are sufficient to reduce any color structure appearing in a pure-gluon tree amplitude to contractions of $f^{abc}$ tensors. See appendix B in ref.~\cite{Johansson:2017srf} for all of these identities. 

Tree-level gluon amplitudes of the so-called $(DF)^2$ theory defined by the Lagrangian (\ref{LagrDF}) have been shown to obey BCJ relations (\ref{2.13}) up to at least eight points. The color-kinematics duality guarantees~\cite{Chiodaroli:2017ngp} that the double copy will give amplitudes that are invariant under space-time diffeomorphisms, i.e.\  amplitudes in a gravitational theory. In terms of tree-level amplitudes, the double copy (\ref{CG_DC}) for conformal supergravity can be expressed using the standard KLT formula, 
\begin{align}
M_{{\rm CSG}} &= \sum_{\tau,\rho} A_{(DF)^2}(1,\tau,n,n{-}1) S[\tau|\rho]_1
A_{\rm SYM}(1,\rho,n{-}1,n) \,,
\label{2.18}
\end{align}
where the external states of the $(DF)^2$ amplitude are restricted to gluons in order to obtain graviton amplitudes in the double copy.

A curious feature of the tree-level amplitudes of $(DF)^2$ theory is that the product between any two polarization vectors $e_i \cdot e_j$ always cancels out through non-trivial identities~\cite{Johansson:2017srf}, e.g. 
\begin{align}
A_{(DF)^2}(1,2,3) &=  -4 (e_1 \cdot k_2)(e_2 \cdot k_3)(e_3\cdot k_1) 
\label{2.91}
\\
A_{(DF)^2}(1,2,3,4) &=    4\frac{ s^2_{12}s^2_{23}}{ s_{13} } 
\Big( \frac{ k_{4} {\cdot} e_1}{s_{23}} - \frac{ k_2 {\cdot} e_1 }{s_{12}} \Big)
\Big( \frac{ k_{1} {\cdot} e_2}{s_{12}} - \frac{ k_3 {\cdot} e_2 }{s_{23}} \Big)
\Big( \frac{ k_{2} {\cdot} e_3}{s_{23}} - \frac{ k_4 {\cdot} e_3 }{s_{12}} \Big)
\Big( \frac{ k_{3} {\cdot} e_4}{s_{12}} - \frac{ k_1 {\cdot} e_4 }{s_{23}} \Big) \, .
\notag
\end{align} 
It implies that when considering dimensional reduction, the scalars that come from the extra-dimensional gluon components will automatically decouple from the theory. 
In the CHY formulation of $(DF)^2$ theory given in~\cite{Azevedo:2017lkz} the $e_i \cdot e_j$ are manifestly absent. The double poles $\sim s_{12}^{-2}$ in (\ref{2.91}) reflect a gluon propagator $\sim k^{-4}$ due to the four-derivative kinetic term in the Lagrangian (\ref{LagrDF}), and the unusual $s_{13}$-channel pole in $A_{(DF)^2}(1,2,3,4) $ that defies the naive expectation from the color ordering, is due to an exchange of the $\vph^{\alpha}$ scalar. This exchange follows from the coupling 
$\sim C^{\alpha ab} \varphi^\alpha F^a_{\mu \nu} F^{b\mu \nu}$ in the Lagrangian (\ref{LagrDF}) along with the structure constants generated from the contraction $C^{\alpha ab} C^{\alpha cd}$, see (\ref{ClebshGordan}).


\section{Bosonic string amplitudes from the double copy}
\label{sec3}

In this section, we will consider the extended gauge theory $(DF)^2+{\rm YM}$ and argue that massless tree amplitudes (\ref{2.8}) and  (\ref{2.12prime}) of the open and closed bosonic string 
amount to the following double-copy constructions
\bea
(\text{open bosonic string})&=& \big(\textrm{Z-theory}\big) \otimes \big((DF)^2+{\rm YM}\big) \,,\label{BS_DC} \\
(\text{closed bosonic string})&=&\big((DF)^2+{\rm YM}\big)\otimes  {\rm sv}\big(\textrm{open bosonic string} \big) \, .
\nn
\eea
Note that the closed bosonic string can be alternatively viewed as a triple copy involving single-valued Z-theory
\beq
(\text{closed bosonic string})= \big((DF)^2+{\rm YM}\big) \otimes  {\rm sv}\text{(Z-theory)}  \otimes \big((DF)^2+{\rm YM}\big)\,,
\label{BS_DCprime}
\eeq
see the discussion around (\ref{CSvsOSprime}) for the analogous triple-copy structure of closed superstrings.

The bosonic field theory, $(DF)^2+{\rm YM}$ entering (\ref{BS_DC}) is defined by the following Lagrangian
\begin{align}
{\cal L}_{(DF)^2 + {\rm YM}}&= \frac{1}{2}(D_{\mu} F^{a\, \mu \nu})^2  - \frac{g}{3} \,   F^3+ \frac{1}{2}(D_{\mu} \varphi^{\alpha})^2  + \frac{g}{2}  \,  C^{\alpha ab}  \varphi^{ \alpha}   F_{\mu \nu}^a F^{b\, \mu \nu }  +  \frac{g}{3!}  \, d^{\alpha \beta \gamma}   \varphi^{ \alpha}  \varphi^{ \beta} \varphi^{ \gamma} \notag \\ 
 & \ \ \ \   -   \frac{1}{2} m^2 (\varphi^{\alpha})^2- \frac{1}{4} m^2 (F^a_{\mu \nu})^2\,,
\label{massdefL}
\end{align}
which augments (\ref{LagrDF}) by mass-dependent terms also considered in ref.~\cite{Johansson:2017srf}. This $(DF)^2+{\rm YM}$ gauge theory smoothly interpolates between the previously introduced $(DF)^2$ theory $(m=0)$ and pure Yang--Mills theory $(m=\infty)$. Its field content comprises a massless gluon, a massive gluon and a massive scalar (see appendix~\ref{appA0} for the propagators). A priori, the mass is a free parameter, but we will find that it has to be related to the inverse string tension $\alpha'$ by
\beq
m^2= -\frac{1}{\alpha'}\,.
\label{massIdentification}
\eeq
Since $m^2$ is negative, it gives rise to tachyonic modes, i.e.\ the massive gluon as well as the massive scalar $\varphi^\alpha$ are tachyonic. In the context of the open and closed bosonic strings, this reflects the appearance of scalar tachyon modes that propagate internally. Note that the Lagrangian (\ref{massdefL}) is normalized such that the limit $m \rightarrow 0$ ($\alpha' \rightarrow \infty$) is well behaved; for a well-behaved  limit $m \rightarrow \infty$ ($\alpha' \rightarrow 0$) we need to multiply the Lagrangian by $m^{-2}=  -\ap$, which explains such overall normalization factors appearing in some of the subsequent amplitudes. 

Based on a few reasonable assumptions, we can argue that the Lagrangian (\ref{massdefL}) is the unique field theory that describes the kinematic factors $B$ in the bosonic-string amplitudes,
\beq
B(1,2,\ldots,n) = -\ap A_{(DF)^2+{\rm YM}}(1,2,\ldots,n) \, \big|_{m^2=-1/\ap}  \ ,
\label{BvsADF}
\eeq 
which results in the double-copy statements given in \eqn{BS_DC}. These assumptions are:
\begin{itemize}
\item[(i)] In the low-energy limit  $\alpha' \rightarrow 0$, the factors  $B$ become YM tree amplitudes.  
\item[(ii)] In the high-energy limit  $\alpha' \rightarrow \infty$, the $B/\alpha'$ are finite and have at most poles $s_{i\ldots j}^{-2}$.  
\item[(iii)] $B$ are partial amplitudes coming from a bosonic gauge theory that obeys the color-kinematics duality in (at least) $D\le 26$ dimensions.
\end{itemize}
The first assumption (i) follows from the field-theory limit of the construction of ref.~\cite{Huang:2016tag}, see the discussion around (\ref{2.11}). Assumption (ii) is an observed property of the explicit examples given in the same paper. Moreover, the general validity of (ii) appears plausible since the limit $\ap \rightarrow \infty$ converts tachyon propagators $\sim \big[ (k_{i}+k_{j})^2 - \frac{1}{\ap} \big]^{-1}$ of the bosonic string to massless propagators $(k_{i}+k_{j})^{-2}$. Since tachyon propagators usually line up with massless propagators in moduli-space integrals\footnote{A typical disk integral in the opening line for the four-point amplitude of the open bosonic string reads 
\beq
2\ap \int^1_0 \frac{ \dd z_2}{z_{12}^2} \prod_{i<j}^3 |z_{ij}|^{2\alpha' s_{ij}} 
= \frac{ (2\ap)^2 s_{13}   }{2\ap s_{12}-1}  \int^1_0 \frac{ \dd z_2}{z_{12}} \prod_{i<j}^3 |z_{ij}|^{2\alpha' s_{ij}} 
= \frac{2\ap s_{13}}{(1-2\ap s_{12})s_{12}} \, F_{(2)} \,^{(2)} \, .
\label{relfactor}
\eeq
We have used integration by parts to relate it to the basis function $F_{(2)} \, ^{(2)} $ in (\ref{2.2}) which in turn defines $B(1,2,3,4)$ via (\ref{2.8}). The relative factor of $\frac{2\ap s_{13}}{(1-2\ap s_{12})s_{12}}$ in (\ref{relfactor}) tends to $-\frac{s_{13}}{s_{12}^2}$ as $\ap \rightarrow \infty$ and thereby introduces a double pole $\sim s_{12}^{-2}$ into $A_{(DF)^2}(1,2,3,4)$.} the limit $\alpha' \rightarrow \infty$ introduces double poles $\sim s^{-2}_{ij}$ as for instance seen in (\ref{2.91}). Additional poles $s_{ij}^{-n}$ of higher order $n\geq 3$ do not occur since the OPE among gluon vertex operators does not introduce any singularities $|z_{ij}|^{2\alpha' k_i \cdot k_j-n}$ with $n\geq 3$ into the worldsheet integrand. Finally, assumption (iii) is consistent with the observation \cite{Huang:2016tag} that the $B$ obeys the well-known relations of gauge-theory partial amplitudes: cyclicity, Kleiss--Kuijf and BCJ relations.   

Under the above three assumptions, both limits $\alpha' \rightarrow 0, \infty$ of $B$ result in gauge-theory amplitudes that have no dimensionful parameter. It is thus a simple task to list all possible operators of a given dimension in the underlying gauge-theory Lagrangians and to constrain unknown couplings using color-kinematics duality. From dimensional analysis, it follows that the limit $\alpha' \rightarrow 0$ is dominated by dimension-four operators (in $D=4$ counting), and $\alpha' \rightarrow \infty$ is dominated by dimension-six operators (in $D=6$ counting). In ref.~\cite{Johansson:2017srf} it was shown that there is a unique color-kinematics satisfying dimension-six Lagrangian that is built out of parity-invariant\footnote{Requiring that the theory is defined in $D\le 26$ dimensions rules out the appearance of a Levi--Civita tensor in the Lagrangian.} operators involving $F_{\mu\nu}$'s and the scalar $\vph^\alpha$, and where the kinetic terms have at most four derivatives.  Once the dimension-four YM term, $(F^a_{\mu \nu})^2$, is introduced, there exists a one-parameter solution to the constraints of color-kinematics duality~\cite{Johansson:2017srf}, this free parameter is the mass $m$ and the resulting Lagrangian the one in \eqn{massdefL}. By mutual consistency of the three assumptions, $m^2$ has to be proportional to $(\alpha')^{-1}$.

Color-stripped amplitudes $A_{(DF)^2+{\rm YM}}$ up to multiplicity eight are straightforward to compute from the Feynman rules of the Lagrangian (\ref{massdefL}). We have confirmed that they reproduce all known kinematic factors $B$ coming from the open-bosonic-string amplitudes (\ref{2.8}), given that the mass is identified with $\alpha'$ as in \eqn{massIdentification}. More precisely we have derived (\ref{2.9}) and (\ref{2.10}) at three and four points as well as the expression
for $B(1,2,3,4,5)$ in the arXiv submission of ref.~\cite{Huang:2016tag} from the gauge-theory Lagrangian (\ref{massdefL}).
These checks and the plausibility of the assumptions (i) to (iii) provide strong evidence for our general conjecture (\ref{BvsADF}) on the field-theory description of $B$. The latter in turn implies the double-copy structures (\ref{BS_DC}) of the massless sector of bosonic open- and closed-string theories.


\section{Heterotic string amplitudes from the double copy}
\label{sec4}

The field-theory origin of the kinematic factors $B$ in the open bosonic string settles the double-copy structure of the gravitational tree amplitudes (\ref{2.12}) of the heterotic string,
\beq
\text{(gravity sector of the heterotic string)}=\big((DF)^2+{\rm YM}\big)\otimes  {\rm sv}\big(\textrm{open superstring} \big) \,.
\label{HET_DC}
\eeq
Given that the amplitude structure is analogous to (\ref{kltlike}) and therefore symmetric under the exchange of the left- and right-moving kinematic factors, one can also rewrite (\ref{2.12}) as
\begin{align}
{\cal M}^{h}_{\rm het} &= \sum_{\tau,\rho} A_{\rm SYM}(1,\tau,n,n{-}1)  \, S[\tau|\rho]_1   \,
{\rm sv} {\cal A}_{\rm bos}(1,\rho,n{-}1,n) \, .
\label{2.12new}
\end{align}
In this representation, the supersymmetries are attributed to the $\ap$-independent
part $ A_{\rm SYM}$ of the KLT-like formula which results in the alternative double-copy formulation
\beq
\text{(gravity sector of the heterotic string)}=\big({\rm SYM}\big)\otimes  {\rm sv}\big(\textrm{open bosonic string} \big) \,.
\label{HET_DC2}
\eeq
Alternatively, (\ref{HET_DC}) and (\ref{HET_DC2}) are connected through the following triple-copy reformulation
\beq
\text{(gravity sector of the heterotic string)} = \big((DF)^2+{\rm YM}\big) \otimes  {\rm sv}\text{(Z-theory)}  \otimes \big({\rm SYM}\big) \,,
\label{triplehet}
\eeq
along the lines of (\ref{CSvsOSprime}) and (\ref{BS_DCprime}).

In this section, we extend this construction to heterotic-string amplitudes involving arbitrary combinations of gauge-multiplet and supergravity-multiplet states. We will spell out a field-theory Lagrangian for the first double-copy component in
\beq
(\text{heterotic string})=\big((DF)^2+{\rm YM}+\phi^3\big)\otimes  {\rm sv}\big(\textrm{open superstring} \big)  
\label{HET_DC3}
\eeq
which is equivalent to the triple copy
\beq
(\text{heterotic string})=\big((DF)^2+{\rm YM}+\phi^3\big)\otimes  {\rm sv}\text{(Z-theory)}  \otimes \big({\rm SYM}\big)  \, .
\label{HET_DC3equiv}
\eeq
%
 
\subsection{The structure of heterotic string amplitudes}
\label{sec40}

Given that the gauge and gravity multiplets of the heterotic string couple via genus-zero worldsheets, 
already the gauge sector exhibits structureful tree-level amplitudes with multiple traces of the color factors. Single-trace amplitudes ${\cal A}_{\rm het}$ can be obtained from the single-valued projection \cite{Stieberger:2014hba} 
\beq
 {\cal A}_{\rm het}(1,2,\ldots,n)=
{\rm sv} {\cal A}_{\text{s}}(1,2,\ldots,n)
 \label{4.1}
\eeq
of the type-I results (\ref{2.1}). Double-trace amplitudes of gauge multiplets in turn have been reduced to linear combinations of their single-trace counterparts (\ref{4.1}) \cite{Schlotterer:2016cxa}. The expansion coefficients in these double-trace results exemplify that multi-trace contributions generically violate uniform transcendentality, and it is an open problem to generalize the explicit relations of ref.~\cite{Schlotterer:2016cxa} to higher numbers of traces. In fact, the Lagrangian given in the next subsection will implicitly reduce any multi-trace sector to linear combinations of (\ref{4.1}).

Mixed amplitudes ${\cal M}^{g+h}_{\rm het}$ involving external gauge-multiplets $g$ and gravity-multiplets $h$ can also be organized according to the trace structure of their color dependence. The single-trace sector with $\leq 3$ gravitons as well as the double-trace sector with $\leq 1$ graviton has been expressed in terms of the gauge amplitudes \cite{Schlotterer:2016cxa}. The amplitude relations of this reference include\footnote{The type-I analogue of (\ref{4.2}) can be found in ref.~\cite{Stieberger:2016lng}, and the field-theory limit of (\ref{4.2}) and (\ref{4.3}) has been computed in the CHY~\cite{Nandan:2016pya} and BCJ~\cite{Chiodaroli:2017ngp} frameworks.}
\begin{align}
{\cal A}_{\text{het}}(1,2,\ldots,n;p)&= \sum_{j=1}^{n-1} (e_{p}\cdot x_{j})\,
{\cal A}_{\rm het}(1,2,\ldots, j,p,j{+}1,\ldots,n) \label{4.2} \\
{\cal A}_{\text{het}}(1,2, \ldots,n; p,q)
&=  
\sum_{1=i\leq j}^{n-1}   (e_{p}\cdot x_{i})\, (e_{q}\cdot x_{j})\, 
{\cal A}_{\text{het}}(1,\ldots,i,p,i{+}1,\ldots, j, q,j{+}1,\ldots, n)  \notag \\
&\! \! \!  \! \! \!  \! \! \!  \! \! \!  \! \! \! \! \! \! \! \! \! \! \! \! \! \! \! \! \! \! \! \! \! \! \! \! \! \! \! \! \! \! \! \! \! \! \! 
- { \big[ (e_{p}  \cdot  e_{q})  - 2\alpha'  (e_{p}  \cdot  q)  (e_{q}  \cdot p)  \big] \over 2 \, \big[1-2\alpha' (p \cdot q) \big]} \sum_{l=1}^{n-1} (p \cdot k_l)  \sum_{1=i\leq j}^l 
{\cal A}_{\text{het}}( 1,2,\ldots, i{-}1,q,i, \ldots,j{-}1,p,j,\dots ,n) 
 \notag 
 \\
&  \! \! \!  \! \! \!  \! \! \!  \! \! \!  \! \! \! \! \! \! \! \! \! \! \! \! \! \! \! \! \! \! \! \! \! \! \! \! \! \! \! \! \! \! \! \! \! \! \! 
  - (e_{q} \cdot p)  \sum_{j=1}^{n-1}
  (e_{p}  \cdot x_{j}) \sum_{i=1}^{j+1} 
\,
{\cal A}_{\text{het}}(1,2,\ldots,i{-}1,q, i, \ldots, j,p, j{+}1,\ldots,n) 
+ (  p \leftrightarrow q) \, ,
\label{4.3}
 \end{align}
with gluon legs $1,2,\ldots,n$, graviton momenta $p,q$, graviton polarizations $e_p,e_q$ as well as region momenta $x_j\equiv k_1 {+}k_2{+}\ldots{+}k_j$. By analogy with the field-theory result \cite{Du:2017gnh}, amplitude relations of this type are expected to exist for all multi-trace sectors of the heterotic string and any number of gluons and gravitons \cite{Stieberger:2014hba, Schlotterer:2016cxa}. This would allow us to reduce the massless tree-level S-matrix of the heterotic string to the same basis of sphere integrals seen in the gravity sector (\ref{2.12}) or the type-II string (\ref{2.5}) and leads us to conjecture that
\begin{align}
{\cal M}^{g+h}_{{\rm het}} &= \sum_{\tau,\rho} A_{(DF)^2+{\rm YM}+\phi^3}(1,\tau,n,n{-}1) \, S[\tau|\rho]_1 \,  {\rm sv} {\cal A}_{\rm s}(1,\rho,n{-}1,n)    
 \label{4.4}
\end{align}
for a suitable choice of non-supersymmetric kinematic factors $A_{(DF)^2+{\rm YM}+\phi^3}(1,\tau,n,n{-}1) $.

The factors of $A_{(DF)^2+{\rm YM}+\phi^3}(1,\tau,n,n{-}1) $ capture the color degrees of freedom of the gluons and the extra polarization vectors of the gravitons such as $e_p$ and $e_q$ in (\ref{4.2}) and (\ref{4.3}). Given that the factors of $A_{\rm SYM}$ within ${\cal A}_{\rm s}$ contribute gauge-multiplet polarizations, the external states in (\ref{4.4}) are gauge and gravity multiplets if $A_{(DF)^2+{\rm YM}+\phi^3}(1,\tau,n,n{-}1)$ involve external scalars $s$ and gluons $g$, respectively. Denoting their appearance in the $i^{\rm th}$ and $j^{\rm th}$ leg by $i_s$ and $j_g$, respectively, the independent three- and four-point instances read
\begin{align}
A_{(DF)^2+{\rm YM}+\phi^3}(1_s,2_s,3_g) &= 2 \delta^{A_1 A_2} (e_3 \cdot k_1) \,,
\label{4.5a} 
\end{align}
as well as
\begin{align}
A_{(DF)^2+{\rm YM}+\phi^3}(1_s,2_s,3_s,4_s) &=
{ -\lambda^2 } \frac{\hat f^{A_1 A_2 B} \hat f^{B A_3 A_4} }{  2  s_{12}}  {-\lambda^2 }\frac{\hat f^{A_2 A_3 B} \hat f^{B A_4 A_1} }{ 2 s_{23}} \notag \\
& \ \ \ \ \  -
 s_{13} \Big\{ 
\frac{ 2\alpha'  \delta^{A_1 A_2} \delta^{A_3 A_4}}{s_{12}(1- 2\ap s_{12}) }+ {\rm cyc}(2,3,4) \Big\}\,, \notag \\
A_{(DF)^2+{\rm YM}+\phi^3}(1_s,2_s,3_s,4_g) &= { i \lambda } \hat f^{A_1 A_2 A_3}  \Big( \frac{ e_4\cdot k_3}{s_{12}} - \frac{ e_4\cdot k_1}{s_{23}} \Big)\,, \label{4.5c} \\
A_{(DF)^2+{\rm YM}+\phi^3}(1_s,2_g,3_s,4_g) &=  \delta^{A_1 A_3} \Big\{  \frac{(e_2 \cdot k_1)(e_4 \cdot k_3)}{s_{12}} + \frac{(e_2 \cdot k_3)(e_4 \cdot k_1)}{s_{14}} \notag \\
& \ \ \ \ \ \ \ \ \  \ \ \ \ + (e_2 \cdot e_4)  + \frac{2\alpha' f_{24}}{1- 2 \ap s_{24}}\Big\}  \, .
\notag
\end{align}
We recall the notation $f_{24}=s_{24}(e_2\cdot e_4) - (e_2 \cdot k_4)(e_4 \cdot k_2)$, and the five-point analogues of (\ref{4.5c}) can be found in appendix \ref{appA1}. The subscript of $A_{(DF)^2+{\rm YM}+\phi^3}$ as well as the notation $A_1,A_2,\ldots$ and $\hat f^{A_1A_2A_3}$ for the adjoint indices and the structure constants will become clear below. Similarly, the coupling $\lambda$ 
controls the relative weight of single-trace and multi-trace contributions\footnote{The coupling $\lambda$ can be inserted into the Kac--Moody currents of the gluon vertex operators, which then obey the following OPE $J^A(z) J^B(0) \sim   \hat f^{ABC} J^C(0)\frac{ \lambda }{\sqrt{2}z} + \frac{ \delta^{AB } }{z^2}$.}, and the conventions for the overall normalization of (\ref{4.4}) are detailed in appendix \ref{appNorm}. Note that the heterotic-string amplitudes (\ref{4.4}) are understood to comprise at least two external gluons, and the purely gravitational cases requiring a different bookkeeping of normalization factors are captured by (\ref{2.12}).

Using a combination of the arguments in ref.~\cite{Broedel:2012rc, Huang:2016tag}, it will be argued in appendix \ref{appA2} that the kinematic factors in (\ref{4.4}) obey BCJ relations
\beq
\sum_{j=2}^{n-1} k_1 \cdot (k_2{+}k_3{+}\ldots{+}k_j) \, A_{(DF)^2+{\rm YM}+\phi^3}(2,3,\ldots,j,1,j{+}1,\ldots,n) =0 \ ,
\label{4.6}
\eeq
for arbitrary number of external gluons and adjoint scalars.

\subsection{Scalar extension of the $(DF)^2$ theory}
\label{sec41}

As a main result on the heterotic-string amplitudes, the kinematic factors $A_{(DF)^2+{\rm YM}+\phi^3}$ in (\ref{4.4}) are conjectured to follow as the color-ordered amplitudes in a field theory involving scalars and vectors with Lagrangian~\cite{Johansson:2017srf}
\begin{align}
{\cal L}_{(DF)^2+{\rm YM}+\phi^3} &=  \frac{1}{2}(D_{\mu} F^{a\, \mu \nu})^2  - \frac{g}{3} \, F^3+ \frac{1}{2}(D_{\mu} \varphi^{\alpha})^2  + \frac{g}{2}  \,  C^{\alpha ab}  \varphi^{ \alpha}   F_{\mu \nu}^a F^{b\, \mu \nu } +  \frac{g}{3!}  \, d^{\alpha \beta \gamma}   \varphi^{ \alpha}  \varphi^{ \beta} \varphi^{ \gamma} \notag \\
& \ \ \     -   \frac{1}{2} m^2 (\varphi^{\alpha})^2  - \frac{1}{4} m^2 (F^a_{\mu \nu})^2+ \frac{1}{2} (D_{\mu} \phi^{aA})^2 + \frac{g}{2} C^{\al ab} \vph^\al  \phi^{aA} \phi^{bA} 
\label{4.7} \\
& \ \ \ + \frac{g\lambda}{3!} f^{abc} \hat f^{ABC} \phi^{aA} \phi^{bB} \phi^{cC} \,,  \notag 
\end{align}
with $F^3$ defined in (\ref{fieldDef}). This augments the Lagrangian (\ref{massdefL}) by massless\footnote{Adding a mass term for the scalars $\phi^{aA}$ allows to address massive gauge bosons coupled to gravity upon double copy with YM amplitudes, see \cite{Chiodaroli:2015rdg} for a double-copy description of spontaneous symmetry breaking.} scalars $\phi^{aA}$ that are charged under a global group $G_{\rm het}$ as well as the local gauge group $G$. We treat it as a single field that carries adjoint gauge-group indices $a,b,c,\ldots$ and global-group adjoint indices $A,B,C,\ldots$, and this allows for a cubic scalar self-interaction with an unconstrained and dimensionless coupling $\lambda$.
Note that the global group $G_{\rm het}$ becomes the gauge group of the heterotic string after the double copy~(\ref{HET_DC3}) is performed, and in the tree-level context of this work, $G_{\rm het}$ can be generalized beyond the rank-sixteen cases realized by compactification of sixteen spacetime dimensions \cite{Gross:1985fr, Gross:1985rr}. As for the $(DF)^2+{\rm YM}$ theory, the Lagrangian (\ref{4.7}) is normalized such that the limit $m \rightarrow 0$ ($\alpha' \rightarrow \infty$) is well behaved. For a well-behaved limit $m \rightarrow \infty$ ($\alpha' \rightarrow 0$) in turn, we need to multiply the Lagrangian by $m^{-2}=  -\ap$, and rescale the bi-adjoint scalar $\phi^{aA}\rightarrow m\phi^{aA}$ and redefine its self-coupling $m \lambda = \lambda'$, which explains why such steps are needed in some of the subsequent formulas. 

The kinematic factors $A_{(DF)^2+{\rm YM}+\phi^3}$ appearing in the massless heterotic-string amplitudes (\ref{4.4}) we claim to arise from the partial tree-level amplitudes of the Lagrangian (\ref{4.7}) after color-decomposing with respect to the local gauge group $G$. As exemplified by (\ref{4.5a}) and (\ref{4.5c}), these color-ordered amplitudes still depend on the degrees of freedom of the global group $G_{\rm het}$ through the appearance of $\delta^{AB}$ and $\hat f^{ABC}$ tensors.

That the identification of the kinematic factor in eq.~(\ref{4.4}) is correct can be argued through similar considerations as in section~\ref{sec3}. By considering the $A_{(DF)^2+{\rm YM}+\phi^3}$ as unknown factors appearing in the massless heterotic-string amplitudes, we assume the following minimal properties (in addition to the properties already considered for the $B$ amplitudes in section~\ref{sec3}),
\begin{itemize}
\item[(i)] In the low-energy limit $ -1/\alpha' = m^2 \rightarrow \infty$ combined with the abelian limit of $G_{\rm het}$ ($\lambda \rightarrow 0$), the factors $A_{(DF)^2+{\rm YM}+\phi^3}$ become tree amplitudes in 26-dimensional YM theory dimensionally reduced to $D=10$, where the scalars arise as internal components of the gluons.
\item[(ii)] In the limit $\lambda \rightarrow \infty$, the multi-trace terms of $A_{(DF)^2+{\rm YM}+\phi^3}$ are suppressed, and their
pure-scalar sector gives trivial $\phi^3$ tree amplitudes. 
\item[(iii)] $A_{(DF)^2+{\rm YM}+\phi^3}$ are partial amplitudes coming from a bosonic gauge theory that obeys the color-kinematics duality in (at least) $D\le 10$ dimensions.
\end{itemize}
The first property comes from the knowledge that heterotic-string amplitudes in the field-theory limit and abelian limit 
become double copies between 26-dimensional YM theory and ten-dimensional SYM \cite{Gross:1985fr, Gross:1985rr}.
The second property comes from considering gluon scattering at tree level in the heterotic string and taking the gauge-group coupling to be large in comparison to the gravitational coupling in units of $\ap$. This limit can be implemented via $\lambda \rightarrow \infty$ which gives single-trace amplitudes of uniform transcendentality, consistent with the single-valued projection (\ref{4.1}) of the open-string amplitudes \cite{Stieberger:2014hba}. A double copy with $\phi^3$ amplitudes on one side is equivalent to the identity operation, and thus property (ii) is consistent with \eqn{HET_DC3}. Assumption (iii) follows from (\ref{4.6}) based on the arguments in appendix \ref{appA2}.

The amplitudes obtained from the Lagrangian~(\ref{4.7}) obey all of the above properties, 
and have the right behaviour in the limits $\alpha' \rightarrow 0$ and $\alpha' \rightarrow \infty$
(see section \ref{sec3} for the analogous $\alpha' \rightarrow 0, \infty$ properties in case of the 
kinematic factors of the bosonic string).
From the analysis in ref.\ \cite{Johansson:2017srf}, ${\cal L}_{(DF)^2+{\rm YM}+\phi^3}$ given in 
(\ref{4.7}) is the unique Lagrangian with these properties.

{\bf More on property (i):}
Property (i) deserves some further explanation as it is not obvious from simply staring at the Lagrangian~(\ref{4.7}). After the rescaling $\phi^{aA}\rightarrow m \phi^{aA}$, the limit $m \rightarrow \infty$ and $\lambda \rightarrow 0$ can be taken by keeping all terms proportional to $m^2$ and dropping the rest. The $\varphi^\alpha$ is now an auxiliary field that can be integrated out, this gives the $\phi^4$ interaction that is typical of dimensionally reduced YM theory: 
\begin{align}
{\cal L}_{{\rm YM}+{\rm scalar}} & ~= ~ \, - \frac{1}{4} (F^a_{\mu \nu})^2 + \frac{1}{2} (D_{\mu} \phi^{aA})^2  -   \frac{1}{2}  (\varphi^{\alpha})^2 + \frac{g}{2} C^{\al ab} \vph^\al  \phi^{aA} \phi^{bA} \,  \notag  \\
 & \longrightarrow - \frac{1}{4} (F^a_{\mu \nu})^2 + \frac{1}{2} (D_{\mu} \phi^{aA})^2   - \frac{g^2}{4} f^{ace}f^{ebd} \phi^{aA} \phi^{bA} \phi^{cB} \phi^{dB} \,,
 \label{YMphi3}
\end{align}
where the overall $m^2$ has been removed by a trivial rescaling. Note that since we obtain dimensionally reduced YM theory, we can simply think of the $\phi^{aA}$ scalars as being the extra-dimensional gluons coming from the 26-dimensional theory. This interpretation works even for finite $m$, but in the limit $m \rightarrow 0$ the interpretation breaks down (as previously mentioned, dimensional reductions of the pure $(DF)^2$ theory yield scalars that automatically decouple~\cite{Johansson:2017srf}). 

{\bf More on property (ii):}
To elaborate on a stronger version of property (ii), we may take the $m \rightarrow \infty$ limit of the Lagrangian~(\ref{4.7}) while keeping the product $\lambda' \equiv m \lambda$ finite (which is now a dimensionful coupling constant). This yields the ${\rm YM}+\phi^3$ Lagrangian introduced in ref.~\cite{Chiodaroli:2014xia},
\begin{align}
{\cal L}_{{\rm YM}+\phi^3} &= - \frac{1}{4} (F^a_{\mu \nu})^2 + \frac{1}{2} (D_{\mu} \phi^{aA})^2 - \frac{g^2}{4} f^{ace}f^{ebd} \phi^{aA} \phi^{bA} \phi^{cB} \phi^{dB} + \frac{g\lambda'}{3!} f^{abc} \hat f^{ABC} \phi^{aA} \phi^{bB} \phi^{cC}  \,, \label{eymLagr}
\end{align}
which gives rise to (possibly supersymmetric) EYM upon double-copy with (S)YM.
The claim (\ref{HET_DC3}) is consistent with the $\ap \rightarrow 0$ limit ($\lambda'= m\lambda$ finite), 
where the heterotic-string amplitude (\ref{4.4}) should reproduce the KLT formula
\begin{align}
M^{g+h}_{{\rm EYM\,SG}} &= \sum_{\tau,\rho} A_{{\rm YM}+\phi^3}(1,\tau,n,n{-}1) S[\tau|\rho]_1 A_{\rm SYM}(1,\rho,n{-}1,n)
 \label{EYMKLT}
\end{align}
for EYM amplitudes. Indeed, the $m \rightarrow \infty$ limit of the kinematic factors $A_{(DF)^2+{\rm YM}+\phi^3}$ in (\ref{4.4}) yields amplitudes of the ${\rm YM}+\phi^3$ theory
\beq
\lim_{m \rightarrow \infty}  m^{n_s-2} A_{(DF)^2+{\rm YM}+\phi^3}(1,2,\ldots,n)\Big|_{\lambda' = m\lambda  \ \te{fixed}}  ~~ =~~ A_{{\rm YM}+\phi^3}(1,2,\ldots,n)  \ ,
\label{FTBDF}
\eeq
where $n_s$ is the number of external scalars, and thus this generalizes the low-energy limit of the pure-gluon amplitudes (\ref{2.11}). Note that the purpose of inserting an overall factor of $ m^{n_s-2}$ and redefinition of $\lambda$ is to make the dimensions agree on the two sides of the equality. (Accordingly, the $\ap \rightarrow 0$ limit of the heterotic amplitude~(\ref{4.4}) needs the analogous rescaling.)

{\bf Consistency as $\ap \rightarrow \infty$:} The compatibility of the double copy (\ref{4.4}) with the
Lagrangian (\ref{4.7}) in the opposite limit $m\rightarrow 0$ can also be seen. Correlation functions of the heterotic string again yield worldsheet integrals whose expansion in a uniform-transcendentality basis of $({\rm sv} F)_\rho{}^\sigma$ involves coefficients $\sim \frac{ 2\ap}{2\ap s_{i\ldots j}-1}$. As explained in section \ref{sec3}, the latter usually line up with massless propagators $s_{i\ldots j}^{-1}$ such that the limit $\ap \rightarrow \infty$ introduces double poles $s_{i\ldots j}^{-2}$ into $A_{(DF)^2+{\rm YM}+\phi^3}$ as expected from the massless limit of (\ref{4.7}).

{\bf Relating $\ap$ to the mass parameter of $(DF)^2+{\rm YM}+\phi^3$ theory:}
At this point, one can also see that the mass parameter $m^2=-(\ap)^{-1}$ of (\ref{4.7}) needs to be interlocked with $\ap$ in the Koba--Nielsen factor of the disk integrals (\ref{2.2}) and their single-valued projection. Our argument is based on the level-matching condition for left- and right-moving degrees of freedom which removes the tachyon from the heterotic-string spectrum. Tachyon propagation in massless four-point amplitudes of the heterotic string is suppressed by the zeros of the Virasoro--Shapiro factor
\beq
{\rm sv} {\cal A}_{\rm s}(1,2,3,4) = \frac{ \Gamma(1+2\ap s_{12}) \Gamma(1+2\ap s_{13}) \Gamma(1+2\ap s_{23}) }{ \Gamma(1-2\ap s_{12}) \Gamma(1-2\ap s_{13}) \Gamma(1-2\ap s_{23})} \, A_{\rm SYM}(1,2,3,4)
\label{level1}
\eeq
at $2s_{ij} = \ap^{-1}$. Like this, the tentative tachyon poles $(1-2\ap s_{12})^{-1}$ of the four-point examples $A_{(DF)^2+{\rm YM}+\phi^3}(1,2,3,4)$ in (\ref{4.5c}) are cancelled by the factor of $\Gamma(1-2\ap s_{12})^{-1}$ within (\ref{level1}). This cancellation crucially depends on having the same value of $\ap$ in the two sides of the double copy (\ref{4.4}). 

These arguments can be repeated in the context of the closed bosonic string: The double copy of the kinematic factors $B$ in the tree amplitudes (\ref{2.12prime}) of the closed bosonic string introduces double poles at $2s_{ij} = - m^{2}$. Once the mass parameter of the underlying $(DF)^2+ {\rm YM}$ theory is adjusted to $\ap$, these double poles conspire with the zeros of the accompanying sphere integrals at $2s_{ij} = \ap^{-1}$. Hence, the field-theory mass parameter must be related to $\ap$ in order to arrive at the kinematic pole structure for tachyon propagation.

The above discussion is tailored to open-string conventions for the normalization of $\ap$. The conventional 
closed-string normalization of $\alpha'$ in (\ref{level1}) and related equations can be attained by replacing $\alpha' \rightarrow \frac{\ap}{4}$ which identifies the mass of the closed-string tachyon to be twice the mass of the open-string tachyon.

{\bf Explicit checks:} Amplitudes up to multiplicity eight can be straightforwardly computed from the Feynman rules of the $(DF)^2+{\rm YM}+\phi^3$ theory (see appendix~\ref{appA0}); we have explicitly checked that these agree with all the corresponding kinematic factors extracted from heterotic string amplitudes up to multiplicity five, as well as the triple-trace color structure\footnote{We are grateful to Sebastian Mizera for sharing the integration-by-parts reduction of the worldsheet integral entering the triple-trace sector of the six-gluon amplitude of the heterotic string.} $\sim \delta^{A_1 A_2}\delta^{A_3 A_4}\delta^{A_5 A_6}$ in the six-point heterotic-string amplitude (\ref{4.4}).


\section{Gauge-gravity amplitude relations  and CSG from  $\ap \rightarrow \infty$}
\label{sec5}

This section is dedicated to field-theory implications of our double-copy representations of string amplitudes.
We discuss some connections between the $\ap \rightarrow \infty$ limit of heterotic-string amplitudes and those of conformal gravity, including its coupling to gauge theories. In this limit, the double-copy structure of the heterotic string is explained to imply
amplitude relations in more exotic examples of gauge-gravity theories, see section \ref{sec44} below.

\subsection{Amplitude relations for conformal supergravity coupled to gauge bosons}
\label{sec42}

As reviewed in section \ref{sec23}, the $\ap \to \infty$ limit of the ($(DF)^2+{\rm YM}$)-theory amplitudes can be double copied with SYM to yield conformal-gravity amplitudes~\cite{Johansson:2017srf} of Berkovits-Witten type~\cite{Berkovits:2004jj} (see also~\cite{Bergshoeff:1982az,deRoo:1991at}). In ref.~\cite{Azevedo:2017lkz}, it was shown that these conformal-gravity amplitudes can also be obtained from the heterotic ambitwistor string. Given the relation between the tensionless limit of usual string theory and ambitwistor strings \cite{Siegel:2015axg,Casali:2016atr,Lee:2017utr}, it is reasonable to expect  the $\ap \rightarrow \infty$ limit of the heterotic-string relations (\ref{4.2}) and (\ref{4.3}) to apply to color-stripped amplitudes $A_{{\rm CSG}+{\rm SYM}}$ in conformal supergravity coupled to gauge bosons.

Indeed, it will be shown in appendix \ref{appB1} through the CHY formalism that this is the case for a small number of external gravitons. While the amplitude coefficients in (\ref{4.2}) for the one-graviton case do not depend on $\ap$ and allow for the immediate translation 
\begin{align}
A_{{\rm CSG}+{\rm SYM}}(1,2,\ldots,n;p)&= \sum_{j=1}^{n-1} (e_{p}\cdot x_{j})\,
A_{\rm SYM}(1,2,\ldots, j,p,j{+}1,\ldots,n) \label{4.2new}  \ ,
\end{align}
the two-graviton relation (\ref{4.3}) explicitly depends on $\ap$ via ${   e_{p}  \cdot  e_{q}  - 2\alpha'  (e_{p}  \cdot  q)  (e_{q}  \cdot p)  \over  1-2\alpha' (p \cdot q) }$  in the second line. The limit $\ap \rightarrow \infty$ reduces this kinematic coefficient to $\frac{(e_p\cdot q)(e_q\cdot p)}{ s_{pq}}$ and therefore gives rise to the amplitude relation
\begin{align}
A_{{\rm CSG}+{\rm SYM}}(1,2, \ldots,n; &p,q)
=  
\sum_{1=i\leq j}^{n-1}   (e_{p}\cdot x_{i})\, (e_{q}\cdot x_{j})\, 
A_{\text{SYM}}(1,\ldots,i,p,i{+}1,\ldots, j, q,j{+}1,\ldots, n)  \notag \\
&\! \! \!  \! \! \!  \! \! \!  \! \! \!  \! \! \! \! \! \! \! \! \! \! \! \! \! \! \! \! \! \! \! \! \! \! \! \! \! \! \! \! \! \! \! \! \! \! \!  \!
- 
\frac{(e_p\cdot q)(e_q\cdot p)}{2 s_{pq}}
\sum_{l=1}^{n-1} (p \cdot k_l)  \sum_{1=i\leq j}^l 
A_{\text{SYM}}( 1,2,\ldots, i{-}1,q,i, \ldots,j{-}1,p,j,\dots ,n) 
\label{4.3new}
 \\
&  \! \! \!  \! \! \!  \! \! \!  \! \! \!  \! \! \! \! \! \! \! \! \! \! \! \! \! \! \! \! \! \! \! \! \! \! \! \! \! \! \! \! \! \! \! \! \! \! \! \!
  - (e_{q} \cdot p)  \sum_{j=1}^{n-1}
  (e_{p}  \cdot x_{j}) \sum_{i=1}^{j+1} 
\,
A_{\text{SYM}}(1,2,\ldots,i{-}1,q, i, \ldots, j,p, j{+}1,\ldots,n) 
+ (  p \leftrightarrow q) \, ,
\notag
 \end{align}
see appendix \ref{appB1} for a detailed derivation from the CHY formalism. 
Note that ${{\rm CSG}+{\rm SYM}}$ is a ten-dimensional generalization of Witten's twistor string theory~\cite{Witten:2003nn} (including the conformal gravity sector that is often discarded). The generalization of (\ref{4.2new}) and (\ref{4.3new}) to three
gravitons, which can also be derived from the $\ap \rightarrow \infty$ limit of the corresponding heterotic-string relations \cite{Schlotterer:2016cxa}, can be found in appendix \ref{appB2}. Further generalizations to arbitrary numbers of gravitons and traces are encoded in the double-copy (\ref{4.4}) along with the Lagrangian in section \ref{sec41}. 

Since the ${{\rm CSG}+{\rm SYM}}$ amplitudes can be obtained from the $\ap \to \infty$ limit of the double copy $((DF)^2 +{\rm YM}+ \phi^3) \otimes {\rm SYM}$ they correspond to a double copy between $(DF)^2 +\phi^3$ and SYM, where the Lagrangian for the $(DF)^2 +\phi^3$ theory is obtained by setting $m=0$ in (\ref{4.7}). From this fact it is clear that (\ref{4.3new}) can be used to straightforwardly identify BCJ numerators of the $(DF)^2+\phi^3$ theory after writing the $A_{\rm SYM}$ factors in a Kleiss--Kuijf basis, in similar fashion to the method introduced in ref.~\cite{Chiodaroli:2017ngp} where corresponding EYM relations employ BCJ numerators of the ${\rm YM}+\phi^3$ theory; see ref.~\cite{Chiodaroli:2017ngp}  for further details.

Further amplitude relations can be obtained for multi-trace sectors of ${{\rm CSG}+{\rm SYM}}$. In appendix \ref{appB3}, double-trace amplitudes $A_{{\rm CSG}+{\rm SYM}}(\{ 1,2,\ldots,r \, | \,r{+}1,\ldots,n \})$ of the gauge multiplet associated with color factors ${\rm Tr}(T^{a_1} T^{a_2} \ldots T^{a_r}) {\rm Tr}(T^{a_{r+1}} \ldots T^{a_n})$ are expressed in terms of (single-trace) gauge-theory amplitudes, e.g.
\begin{align}
A_{{\rm CSG}+{\rm SYM}}(\{ 1,2\, |\, 3,4 \}) &=  \frac{ s_{14} }{s_{12}}  \, A_{\rm SYM}(1,2,3,4)\,,  \\
A_{{\rm CSG}+{\rm SYM}}(\{ 1,2,3\, |\, 4,5 \}) &= \frac{ 1}{ s_{45}}  \, \big[s_{15} A_{\rm SYM}(1,2,3,4,5) - s_{25} A_{\rm SYM}(2,1,3,4,5) \big]  \,.
\end{align}
The multiparticle generalization of these examples and the CHY integrand associated with the double-trace sector can be found in appendix \ref{appB3}.

\subsection{More exotic examples of gauge-gravity theories}
\label{sec44}

In this subsection, we will investigate more exotic gauge-gravity theories that can be obtained by double-copying $(DF)^2$ with massless gauge theories coupled to a bi-adjoint scalar. These additional gauge-gravity theories will be argued to share the amplitude relations of either EYM or ${\rm CSG}+{\rm SYM}$.

As shown in  \cite{Azevedo:2017lkz}, the double copy of $(DF)^2$ with itself produces the amplitudes of $R^3$ (six-derivative) gravity coming from bosonic ambitwistor strings. If one instead considers the double copy of $(DF)^2$ with $(DF)^2 + \phi^3$,  one then obtains the amplitudes of $R^3$ gravity coupled to $(DF)^2$ gauge bosons. Interestingly, given the similarity of the CHY integrands of both theories \cite{Azevedo:2017lkz}, it is easy to show that the amplitude relations in this section and appendices  \ref{appB2} and \ref{appB3} can be carried over to this $R^3 + (DF)^2$ theory. For this purpose, the amplitude subscripts in (\ref{4.2new}), (\ref{4.3new}) and their generalizations need to be adjusted according to ${\rm CSG}+{\rm SYM} \to R^3 + (DF)^2$ and ${\rm SYM} \to (DF)^2$, respectively. Since we show in this paper that the amplitude relations for ${\rm CSG}+{\rm SYM}$ are related to an $\ap \to \infty$ limit, it is reasonable to suspect the same is true for $R^3 + (DF)^2$. It would be interesting to identify the finite-$\ap$ version of these relations.

In table \ref{EYMoverview} below we give an overview of possible double-copy constructions involving gauge theories coupled to a bi-adjoint scalar. While the entries of $R^3$ gravity and its couplings to the $(DF)^2$ theory have just been discussed, it remains to introduce one last gauge-gravity theory: The entry ``${\rm CG}+(DF)^2$'' in table \ref{EYMoverview} refers to conformal gravity coupled to $(DF)^2$ gauge bosons.

\begin{table}[h]
\beq \! \! \!
{\setstretch{1.75} 
\begin{array}{c||c|c|c|c}
\otimes &\te{SYM} &(DF)^2 &\te{YM}+\phi^3 &(DF)^2 + \phi^3 \\\hline \hline
\te{SYM} \ &\ \te{supergravity} \ \, &\te{CSG} &\te{EYM\ SG} &\te{CSG} + \te{SYM} \\
(DF)^2 \ &\te{CSG} &\ R^3 \ \te{gravity} \  \, &\ \te{CG} + (DF)^2  \  \,&\ R^3 \ \te{gravity} + (DF)^2 \  \,
\end{array}} \nonumber
\eeq
\caption{Overview of the double-copy constructions discussed here involving pure gauge theories, or gauge theories coupled to a bi-adjoint scalar.}
\label{EYMoverview}
\end{table}

The fact that the amplitudes coming from the theories in the last column of the table satisfy the same type of relations suggests that the same is true for the next-to-last column. Indeed, amplitudes of ${\rm CG}+(DF)^2$ will be argued to satisfy the same relations as EYM amplitudes: For one graviton, one has
\begin{align}
A_{{\rm CG}+(DF)^2}(1,2,\ldots,n;p)&= \sum_{j=1}^{n-1} (e_{p}\cdot x_{j})\,
A_{(DF)^2}(1,2,\ldots, j,p,j{+}1,\ldots,n) \label{5.6new}  \ ,
\end{align}
while for two gravitons, one inherits the structure of (\ref{4.3}) in the $\ap \rightarrow 0$ limit,
\begin{align}
A_{{\rm CG}+(DF)^2}(1,2, \ldots,n; p,q)
&=  
\sum_{1=i\leq j}^{n-1}   (e_{p}\cdot x_{i})\, (e_{q}\cdot x_{j})\, 
A_{(DF)^2}(1,\ldots,i,p,i{+}1,\ldots, j, q,j{+}1,\ldots, n)  \notag \\
&\! \! \!  \! \! \!  \! \! \!  \! \! \!  \! \! \! \! \! \! \! \! \! \! \! \! \! \! \! \! \! \! \! \! \! \! \! \! \! \! \! \! \! \! \! \! \! \! \!  \!
- 
\frac{1}{2}(e_{p}  \cdot  e_{q})
\sum_{l=1}^{n-1} (p \cdot k_l)  \sum_{1=i\leq j}^l 
A_{(DF)^2}( 1,2,\ldots, i{-}1,q,i, \ldots,j{-}1,p,j,\dots ,n) 
\label{5.7new}
 \\
&  \! \! \!  \! \! \!  \! \! \!  \! \! \!  \! \! \! \! \! \! \! \! \! \! \! \! \! \! \! \! \! \! \! \! \! \! \! \! \! \! \! \! \! \! \! \! \! \! \! \!
  - (e_{q} \cdot p)  \sum_{j=1}^{n-1}
  (e_{p}  \cdot x_{j}) \sum_{i=1}^{j+1} 
\,
A_{(DF)^2}(1,2,\ldots,i{-}1,q, i, \ldots, j,p, j{+}1,\ldots,n) 
+ (  p \leftrightarrow q) \, .
\notag
 \end{align}
Similarly, for a higher number of gauge bosons and gravitons as well as for the multi-trace sectors, one can export the amplitude relations of EYM \cite{Stieberger:2016lng, Nandan:2016pya, Teng:2017tbo, Du:2017gnh} by adjusting $A_{{\rm EYM}} \rightarrow A_{{\rm CG}+(DF)^2}$ and $A_{{\rm SYM}} \rightarrow A_{(DF)^2}$. In appendix \ref{appB7}, we introduce the CHY integrand for the ${\rm CG}+(DF)^2$ theory and justify the agreement of its
amplitude relations with those of EYM.

\subsection{Conformal-supergravity amplitudes and twisted heterotic strings}
\label{sec53}

As pointed out by Huang, Siegel and Yuan~\cite{Huang:2016bdd}, type-II amplitudes (\ref{2.5}) can be converted into those of supergravity by flipping the sign of $\ap$ in one of the open-string constituents of their KLT representation. Introducing the notation
\beq
\overline{ {\cal A}}_{\rm s}(\pi(1,2,\ldots,n)) \equiv {\cal A}_{\rm s}(\pi(1,2,\ldots,n)) \big|_{\ap \rightarrow -\ap}
\label{5.1}
\eeq
for the formal $\ap \rightarrow -\ap$ operation, this statement is
\begin{align}
M_{{\rm SG}} &=  \sum_{\pi,\sigma} {\cal A}_{{\rm s}}(1,\pi(2,3,\ldots,n{-}2),n,n{-}1) {\cal S}_{\ap}[\pi|\sigma]_1
\overline{ {\cal A}}_{\rm s}(1,\sigma(2,3,\ldots,n{-}2),n{-}1,n) \ ,
\label{5.2}
\end{align}
where the $\ap$-dressed KLT matrix ${\cal S}_{\ap}[\pi|\sigma]_1$ is related to its field-theory incarnation (\ref{2.6}) by promoting the momentum dependence to $(k_1{+}k_B)\cdot k_j \rightarrow \frac{\sin(2 \pi \ap (k_1{+}k_B)\cdot k_j )}{2\pi \alpha'}$ with ${\cal S}_{\ap}[2|2]_1 = \frac{ \sin(2 \pi \ap s_{12}) }{2\pi \ap}$.   By the organization of the type-II amplitudes in ref.~\cite{Schlotterer:2012ny}, the entire $\ap$-dependence has been explained \cite{Huang:2016bdd} to drop out from (\ref{5.2}) and only the supergravity amplitude remains. As pointed out in \cite{Mizera:2017cqs}, both (\ref{5.2}) and the conventional string-theory KLT relations \cite{Kawai:1985xq} may be understood as a consequence of the so-called twisted period relations \cite{intersection}, a result in intersection theory.

Since this cancellation of all MZVs can be traced back to properties of the sphere integrals, the analogous combination with one of ${\cal A}_{\rm s}$ replaced by ${\cal A}_{\rm bos}$ reduces to the kinematic factors
\begin{align}
&\sum_{\pi,\sigma} {\cal A}_{{\rm bos}}(1,\pi(2,3,\ldots,n{-}2),n,n{-}1) {\cal S}_{\ap}[\pi|\sigma]_1
\overline{ {\cal A}}_{\rm s}(1,\sigma(2,3,\ldots,n{-}2),n{-}1,n)  \label{5.3a} \\
&\ \ = \sum_{\tau,\rho} B(1,\tau(2,3,\ldots,n{-}2),n,n{-}1) S[\tau|\rho]_1
A_{\rm SYM}(1,\rho(2,3,\ldots,n{-}2),n{-}1,n)\,, \notag
\end{align}
i.e.\ to the zero-transcendentality piece of the gravitational heterotic-string amplitude (\ref{2.12}).

Following the discussion of section \ref{sec3}, the kinematic factors $B$ of the bosonic string degenerate to the double-copy constituent $A_{(DF)^2}$ of conformal supergravity when taking the limit $\ap \rightarrow \infty$. Then, the right hand side of (\ref{5.3a}) reduces to the KLT representation (\ref{2.18}) of the tree amplitudes of conformal supergravity \cite{Johansson:2017srf}, i.e.\ we are led to the following new representation:
\begin{align}
M_{{\rm CSG}} =  -\lim_{\ap \rightarrow \infty} \frac{1}{\ap}
\sum_{\pi,\sigma} {\cal A}_{{\rm bos}}(1,\pi(2,\ldots,n{-}2),n,n{-}1) {\cal S}_{\ap}[\pi|\sigma]_1
\overline{ {\cal A}}_{\rm s}(1,\sigma(2,\ldots,n{-}2),n{-}1,n)\,.
\label{5.3}
\end{align}
Since the tensionless limit of these ``twisted" heterotic string amplitudes gives the same result as the ambitwistor strings, this is another way of seeing that the amplitudes coming from the gravity sector of the heterotic ambitwistor string correspond to conformal-supergravity amplitudes. 

Likewise, we can compactify the bosonic-string factor of (\ref{5.3}) to the same sixteen-dimensional internal manifolds as used for the heterotic string, giving gauge groups $SO(32)$ or $E_8 \times E_8$ (or more general gauge groups at tree level) and tree amplitudes
\beq
 {\cal A}^{\rm comp}_{\rm bos}(1,\pi(2,3,\ldots,n{-}2),n{-}1,n) = \sum_{\rho}  F_\pi{}^\rho A_{(DF)^2+{\rm YM}+\phi^3}(1,\rho(2,3,\ldots,n{-}2),n{-}1,n) \, .
 \label{A.X1}
\eeq
Then, taking the high-energy limit $\ap \rightarrow \infty$ of the ``twisted" heterotic string yields conformal supergravity coupled to SYM,\footnote{In contrast to (\ref{FTBDF}), the $\ap \rightarrow \infty$ limit in (\ref{5.11}) is performed at fixed $\lambda$ rather than at fixed $\lambda' = m \lambda$. By the relative normalization (\ref{BvsADF}) of $B$ and $A_{(DF)^2 + {\rm YM}}$, there is no analogue of the factor $\alpha'^{-1}$ in (\ref{5.3}).}
\begin{align}
M_{{\rm CSG+SYM}} = \lim_{\ap \rightarrow \infty} 
\sum_{\pi,\sigma} {\cal A}^{\rm comp}_{{\rm bos}}(1,\pi,n,n{-}1)  \Big|_{\lambda  \ \te{fixed}}{\cal S}_{\ap} [\pi|\sigma]_1
\overline{ {\cal A}}_{\rm s}(1,\sigma,n{-}1,n)\,,
\label{5.11}
\end{align}
and reproduces the amplitude relations in section \ref{sec42} upon color decomposition.
When translating these ten-dimensional amplitudes (\ref{5.11}) to four dimensions, we obtain Witten's twistor string theory~\cite{Witten:2003nn}, and the corresponding scattering equations present in the Roiban--Spradlin--Volovich (RSV) formula~\cite{Roiban:2004yf} are now a natural consequence of the $\ap \rightarrow \infty$ limit, as originally considered by Gross and Mende~\cite{Gross:1987ar}. 


\section{Summary and Outlook}
\label{sec6}

In this work, we have proposed field-theory double-copy structures for all massless tree-level amplitudes of bosonic and heterotic strings.
One side of the double copy comprises the same Z-theory amplitudes or moduli-space integrals over punctured disk and sphere worldsheets as seen in their superstring counterparts. The additional, non-supersymmetric double-copy components specific to bosonic and heterotic strings are conjectured to descend from massive gauge theories whose Lagrangians are spelt out in the main text. These gauge-theory Lagrangians are combinations of dimension-four and dimension-six operators whose relative scaling is controlled by the mass parameter $\ap= - m^{-2}$. The dimension-six part which dominates in the $\ap \rightarrow \infty$ limit is known as the $(DF)^2$ gauge theory in the double-copy construction of conformal supergravity \cite{Johansson:2017srf}.

Our results give particularly striking examples for the ubiquity of field-theory structures in string amplitudes and universal properties across different string theories. Apart from the universal $\ap$-dependent moduli-space integrals, tree-level amplitudes in different string theories are characterized by gauge theories which specify the kinematic factors. Properties of the moduli-space integrals require these field theories to obey the duality between color and kinematics in the same way as the Z-theory amplitudes do. Given that the respective string-theory context fixes the mass dimensions and the particle content of the field-theory operators, the $\ap \rightarrow 0,\infty$ behaviour and the color-kinematics duality uniquely specify the extensions of the $(DF)^2$ theory relevant to bosonic and heterotic strings. 

We also showed that relations analogous to those of EYM hold for the amplitudes involving gauge and gravity multiplets in conformal (super-)gravity, and that they can be obtained by taking the $\ap \to \infty$ limit of the corresponding heterotic-string amplitudes. Since the theory of gluons coupled to conformal gravity is precisely the one coming from the heterotic ambitwistor string, our results add to the claims that the latter is more naturally associated with the tensionless limit of ordinary strings.

While the results of this work only concern massless external states, the organization of the moduli-space integrals into
$(n{-}3)!\times (n{-}3)!$ bases should extend to scattering of massive string resonances \cite{Huang:2016tag}. It is 
conceivable that the accompanying kinematic factors depending on the higher-spin polarizations can also be traced back
to field theories that generalize the $(DF)^2+{\rm YM}+\phi^3$ Lagrangian.

Another important line of follow-up research concerns loop amplitudes. For massless one-loop amplitudes of maximally supersymmetric open strings, a double-copy structure resembling matrix elements of gravitational $R^4$ operators has been recently proposed in ref.~\cite{Mafra:2017ioj}. The (generalized) elliptic functions of the worldsheet punctures in the reference are tailored to maximal supersymmetry, but the results of ref.~\cite{Berg:2016wux} on one-loop open-string amplitudes in orbifold compactifications might pave the way to generalizations with reduced supersymmetry. It would be interesting to investigate similar double-copy structures in one-loop amplitudes of bosonic and heterotic strings as well as higher-loop generalizations.

While the double-copy structure of string amplitudes is perturbative in the current context, the fact that kinematic factors of type I, IIA/B and heterotic theories uniformly arise from a field-theory perspective, with a simple Lagrangian interpretation, suggests a connection to non-perturbative string dualities. Understanding the role of the $(DF)^2+{\rm YM}+\phi^3$ gauge theory and Z-theory at loop level may shed new light on these dualities, such as the heterotic/type-I duality \cite{Polchinski:1995df}, or even M-theory \cite{Witten:1995ex}.


\section*{Acknowledgements} 

We thank Yu-tin Huang, Arthur Lipstein, Gustav Mogull, Fei Teng and Congkao Wen for many useful discussions 
and collaboration on closely related topics. We also thank Tim Adamo, Johannes Broedel and Arkady 
Tseytlin for enlightening discussions on conformal gravity, and Murat G\"{u}naydin and Radu Roiban for insightful conversations 
related to compactification. We thank Freddy Cachazo and Sebastian Mizera for helpful comments on the final draft.  The research of T.A., M.C. and H.J. is supported by the Knut and Alice Wallenberg Foundation under grant KAW 2013.0235, the Ragnar S\"{o}derberg Foundation under grant S1/16, and the Swedish Research Council under grant 621-2014-5722. The research of O.S.\ and M.C.\ was supported in part by the Munich Institute for Astro- and Particle Physics (MIAPP) of the DFG cluster of excellence ``Origin and Structure of the Universe''. The research of O.S.\ was supported in part by Perimeter Institute for Theoretical Physics. Research at Perimeter Institute is supported by the Government of Canada through the Department of Innovation, Science and Economic Development Canada and by the Province of Ontario through the Ministry of Research, Innovation and Science.


\appendix

\section{Overall normalizations of color/coupling-stripped amplitudes}
\label{appNorm}

This appendix summarizes our normalization conventions for the color-ordered and/or 
coupling-stripped string- and field-theory amplitudes given in the main text. With the shorthand
\beq
C(1,2,3,\ldots,n)={\rm Tr}\big(T^{a_1} T^{a_{2}} T^{a_{3}} \cdots T^{a_{n}}\big)
\eeq
for Chan--Paton color factors, the color- and coupling-dressed string amplitudes are given by
\begin{align}
{\cal A}_{\rm open\,bos}&= g^{n-2} \sum_{\sigma \in S_{n-1}}{\cal A}_{\rm bos}\big(1,\sigma(2,3,\ldots,n)\big) \,C\big(1,\sigma(2,3,\ldots,n)\big)\,, \nn \\
{\cal A}_{\rm type\,I}&= g^{n-2} \sum_{\sigma \in S_{n-1}}{\cal A}_{\rm s}\big(1,\sigma(2,3,\ldots,n)\big) \, 
C\big(1,\sigma(2,3,\ldots,n)\big)\,, \nn  \\
{\cal M}_{\rm heterotic}^{g+h} \Big|_{{\rm single}\atop{\rm trace}} &= g^{n_g-2} \Big(\frac{\kappa}{2}\Big)^{n_h}  \sum_{\sigma \in S_{n_g-1}}{\cal A}_{\rm het}\big(1,\sigma(2,3,\ldots,n_g);p,q,\ldots \big) \,C\big(1,\sigma(2,3,\ldots,n_g)\big)\,,  \nn \\
{\cal M}_{\rm heterotic}^{g+h}&=\Big(\frac{\kappa}{2}\Big)^{n-2} {\cal M}_{\rm het}^{g+h}\,,\nn \\
{\cal M}_{\rm closed\,bos}&=\Big(\frac{\kappa}{2}\Big)^{n-2} {\cal M}_{\rm bos}\,,  \nn \\
{\cal M}_{\rm type\,II}&=\Big(\frac{\kappa}{2}\Big)^{n-2} {\cal M}_{\rm II}\,, 
\end{align}
where $n_g$ and $n_h$ denote the numbers of external gluons and gravitons, respectively, with total 
$n=n_g+n_h$. The coupling constants $\kappa$ and $g$ appear in the ten-dimensional low-energy effective action as $G^{-1/2}{\cal L}_{\rm GR} \sim \frac{4}{\kappa^2}R$ and $G^{-1/2}{\cal L}_{\rm YM}  \sim\frac{1}{4}(F^a_{\mu\nu})^2$, where the metric and the field strength are expanded as $G_{\mu \nu} =\eta_{\mu \nu}+ \kappa h_{\mu \nu}$ and  $F_{\mu \nu}^a = \partial_{\mu} A_{\nu}^a-\partial_{\nu} A_{\mu}^a + g  f^{abc} A_{\mu}^b A_{\nu}^c$. Note that the color-stripped heterotic single-trace amplitude ${\cal A}_{\rm het}$ has a slightly different overall coupling-dependent factor relative to the color-dressed amplitude ${\cal M}_{\rm het}^{g+h}$; the relative factor is $(2g/\kappa)^{n_g-2}$ which is proportional to $\lambda^{n_g-2}$.

The corresponding field-theory amplitudes are normalized in analogous manner, e.g.
\begin{align}
{\cal A}_{\rm gauge}&= g^{n-2} \sum_{\sigma \in S_{n-1}}A_{\rm gauge}\big(1,\sigma(2,3,\ldots,n)\big) \,C\big(1,\sigma(2,3,\ldots,n)\big)\,,\nn \\
{\cal M}_{\rm gravity}&=\Big(\frac{\kappa}{2}\Big)^{n-2} M_{\rm gravity}\,,
\end{align}
where the ``gauge'' subscript stands for YM, SYM or $(DF)^2$ theory, including the deformations of the $(DF)^2$ theory, and ``gravity'' stands for Einstein, EYM, conformal supergravity, or deformations thereof.

\section{Single-valued MZVs in the $\ap$-expansion}
\label{appD}

For tree amplitudes (\ref{2.2}) and (\ref{2.5}) of the open and closed superstring, the $\ap$-expansion gives rise to MZVs (\ref{2.3}) in a uniformly transcendental pattern \cite{Aomoto, Terasoma, Brown:2009qja, Stieberger:2009rr, Schlotterer:2012ny}: The $w^{\rm th}$ order in $\ap$ is exclusively accompanied by MZVs (\ref{2.3}) of weight $w=n_1+n_2+\ldots+n_r$. Moreover, the $(n{-}3)! \times (n{-}3)!$ matrix form of (\ref{2.2}) is suitable to relate coefficients of different MZVs that are conjecturally independent over $\mathbb Q$: Let $P_w$ and $M_w$ denote $(n{-}3)! \times (n{-}3)!$ matrices whose coefficients are degree-$w$ polynomials in $\ap s_{ij}$ with rational coefficients, then (upon suppressing the indices $\pi,\rho \in S_{n-3}$ in (\ref{2.2})) \cite{Schlotterer:2012ny}
\begin{align}
F &= 1 + \zeta_2 P_2 + \zeta_3 M_3 + \zeta^2_2 P_4+ \zeta_5 M_5+ \zeta_3\zeta_2 P_2M_3 + \zeta^3_2 P_6+ \tfrac{1}{2} \zeta_3^2 M_3^2 + \zeta_7 M_7 + \zeta_5 \zeta_2 P_2 M_5  \notag \\
& \ \ \ \ \ \ \   + \zeta_3 \zeta_2^2 P_4 M_3 + \zeta_2^4 P_8 + \tfrac{1}{2} \zeta_3^2 \zeta_2 P_2 M_3^2 + \zeta_3 \zeta_5 M_5 M_3 + \tfrac{1}{5} \zeta_{3,5} [M_5,M_3] + {\cal O}(\ap^9) \, .
\label{2.4}
\end{align}
For instance, the coefficient of $\zeta_2\zeta_3$ is a matrix product $P_2 M_3$ involving the coefficients $P_2$ and $M_3$ of $\zeta_2$ and $\zeta_3$, and similarly, $ \zeta_{3,5}$ is accompanied by a matrix commutator of $M_3$ and $M_5$ (see \cite{Schlotterer:2012ny} for the continuation to higher weights and explanation of the rational prefactor~$\frac{1}{5}$). The state-of-the-art-methods\footnote{An earlier all-multiplicity method is based on polylogarithm manipulations \cite{Zfunctions}. Moreover, five-point expansions have been performed to arbitrary orders in $\ap$ by exploiting the connection with hypergeometric function \cite{Boels:2013jua, Puhlfuerst:2015gta} (also see \cite{Oprisa:2005wu, Stieberger:2006te, Stieberger:2007jv} for $(N{\leq}7)$-point results at certain orders in $\ap$).} to obtain the explicit form of the $P_w$ and $M_w$ at $n$ points involve the Drinfeld associator \cite{Broedel:2013aza} (also see \cite{Drummond:2013vz}) or a Berends--Giele recursion for Z-theory amplitudes \cite{Mafra:2016mcc}. Results up to and including seven points are available for download via \cite{wwwMZV}.

Once the closed-string tree amplitude is cast into the form (\ref{2.5}), the matrix ${\rm sv} F$ can be obtained by imposing certain selection rules on the MZVs in the analogous open-string expansion (\ref{2.4}) which are obscured by the string-theory KLT formula \cite{Kawai:1985xq}. These selection rules have been pinpointed to arbitrary weights \cite{Schlotterer:2012ny} (see \cite{Stieberger:2009rr} for earlier work on weights $\leq 8$) and were later on identified \cite{Stieberger:2013wea} with the single-valued projection \cite{Schnetz:2013hqa, Brown:2013gia}. As exemplified by the leading orders
\begin{align}
{\rm sv} F &= 1 + 2 \zeta_3 M_3 + 2\zeta_5 M_5+2 \zeta_3^2 M_3^2 +2 \zeta_7 M_7 +2 \zeta_3 \zeta_5 \{ M_5,M_3  \}+ {\cal O}(\ap^9) \ ,
\label{2.7}
\end{align}
the closed-string $\ap$-expansion follows from (\ref{2.4}) by acting with the single-valued projection (\ref{svmap}) \cite{Schnetz:2013hqa, Brown:2013gia} on the MZVs, see the references for ${\rm sv} (\zeta_{n_1,n_2,\ldots,n_r})$ at higher weight and depth.

\section{More on $(DF)^2+{\rm YM}+\phi^3$ amplitudes}
\label{appA}

\subsection{Feynman rules}
\label{appA0}

As a gentle warning to the reader, we note that the gauge-theory Lagrangians throughout the paper are given using a Lorentzian metric of mostly-minus signature.  Amplitudes obtained from standard Feynman-rule calculations in this metric are then compared with the string-theory results after flipping the metric signature. Consequently, all amplitudes found in the paper use a mostly-plus metric. 

We now collect the Feynman rules employed for calculations in the mass-deformed scalar-coupled $(DF)^2+{\rm YM}+\phi^3$ theory with Lagrangian (\ref{4.7}).
In mostly-minus signature, the propagators for the gluon $A^{a}_\mu$, scalars $\varphi^\alpha$ and $\phi^{aA}$ are as follows: 
\begin{eqnarray}
\begin{array}{c}\includegraphics[width=0.3\textwidth]{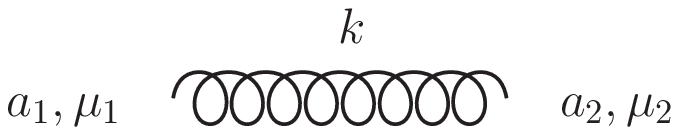} \end{array} &=& \ \
  i {\eta^{\mu_1 \mu_2} \delta^{a_1 a_2} \over k^2 (k^2-m^2)} 
 \\
\begin{array}{c}\includegraphics[width=0.3\textwidth]{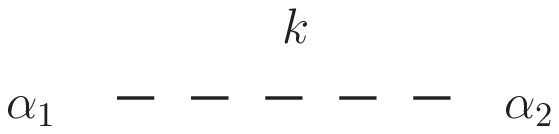} \end{array} &=& \ \
 i { \delta^{\alpha_1 \alpha_2} \over k^2 - m^2} 
 \\
\begin{array}{c}\includegraphics[width=0.3\textwidth]{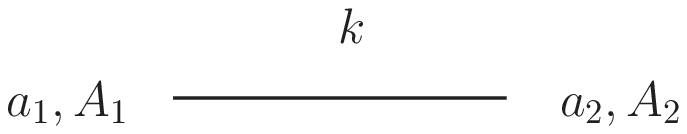} \end{array} &=& \ \
i { \delta^{a_1 a_2} \delta^{A_1 A_2}  \over k^2}  
\end{eqnarray}
The three- and four-vertices in mostly-minus signature are:
\begin{eqnarray}
\begin{array}{c}\includegraphics[width=0.3\textwidth]{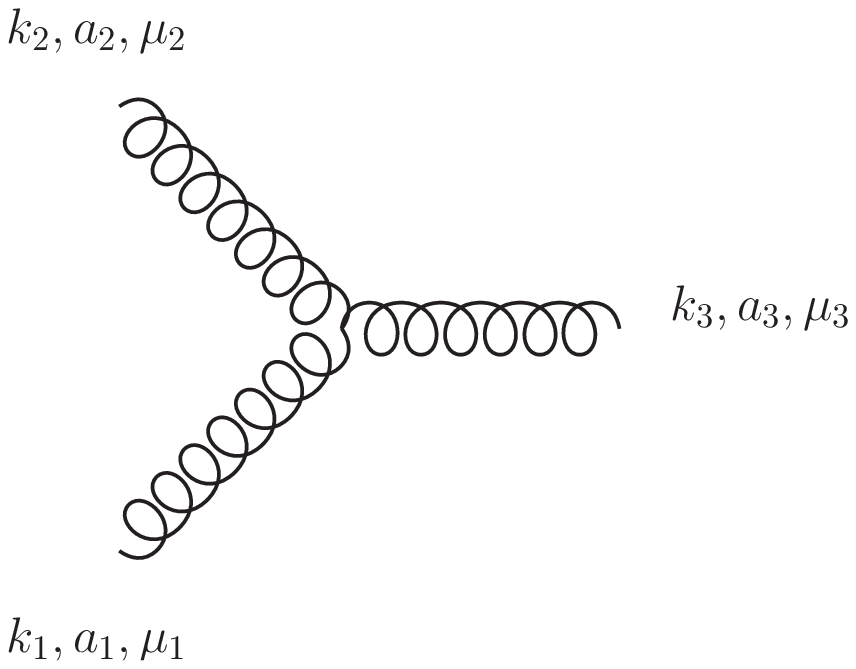} \end{array} &=&
\begin{array}{l} \\  g f^{a_1a_2a_3} \Big\{ (k_1^2+k_2^2 -m^2) \eta^{\mu_1 \mu_2} k_1^{\mu_3} - k_1^{\mu_1}k_2^{\mu_2}k_1^{\mu_3}  \\
-2 k_2^{\mu_1}k_2^{\mu_2}k_1^{\mu_3} +{2\over 3} k_3^{\mu_1}k_1^{\mu_2}k_2^{\mu_3} \Big\} + \text{Perms}(1,2,3) 
\end{array} \\
\begin{array}{c}\includegraphics[width=0.3\textwidth]{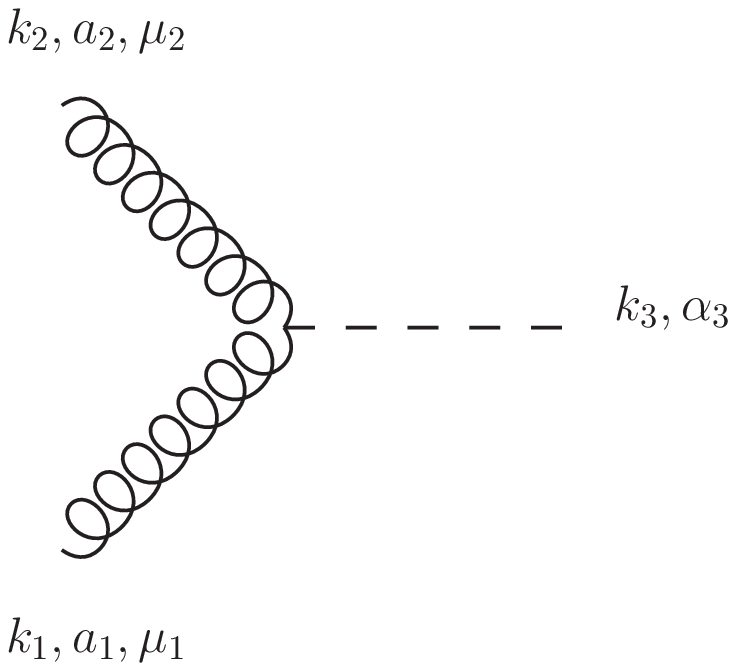} \end{array} &=&
\begin{array}{l}  2 i g C^{\alpha_3 a_1a_2}  (k_2^{\mu_1}k_1^{\mu_2} - k_1 \cdot k_2 \eta^{\mu_1\mu_2}) 
\end{array} \\
\begin{array}{c}\includegraphics[width=0.3\textwidth]{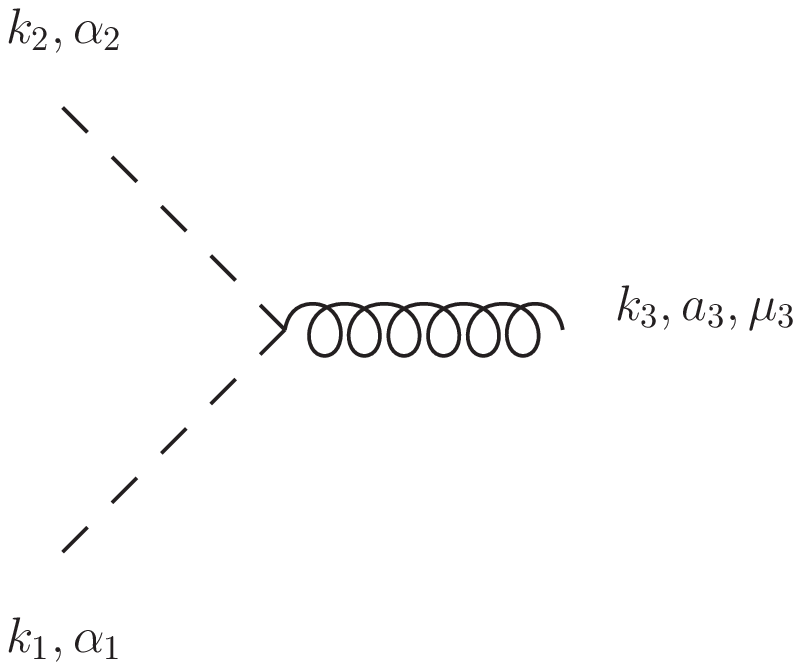} \end{array} &=&
\begin{array}{l}  i g  T^{a_3 \alpha_1 \alpha_2}_R (k_{1}^{\mu_3}- k_{2}^{\mu_3}) 
\end{array} \\
\begin{array}{c}\includegraphics[width=0.3\textwidth]{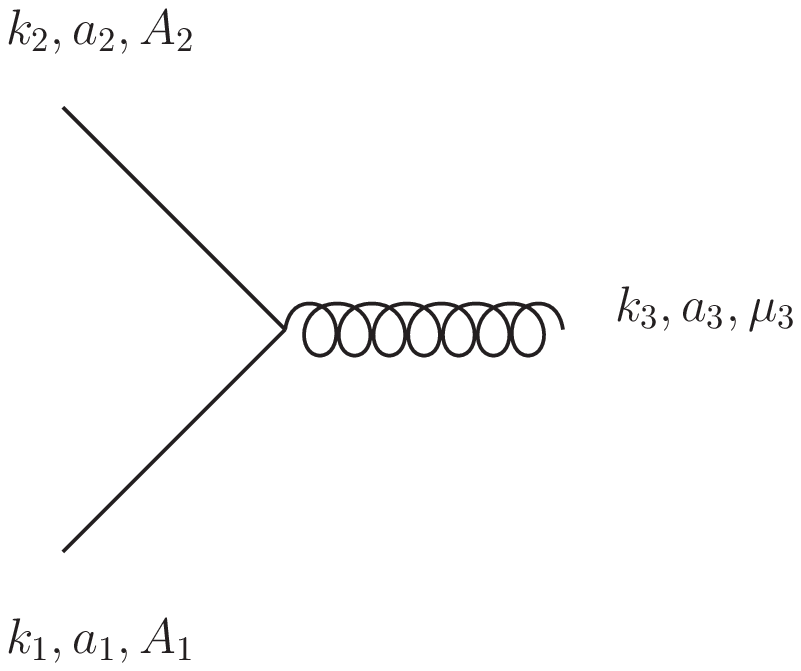} \end{array} &=&
\begin{array}{l}  g f^{a_1a_2a_3} \delta^{A_1 A_2}  \big( k_1^{\mu_3}-k_2^{\mu_3} \big) 
\end{array} \\
\begin{array}{c}\includegraphics[width=0.3\textwidth]{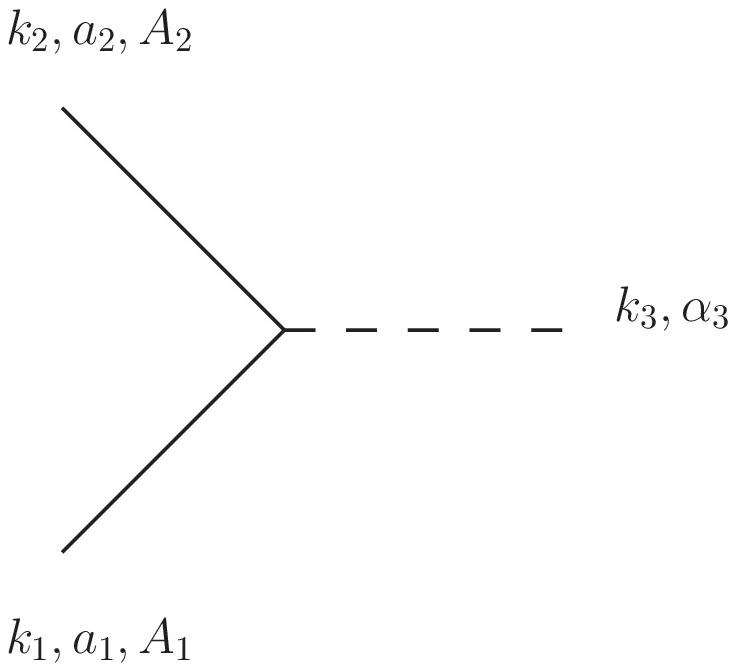} \end{array} &=&
\begin{array}{l} i g C^{\alpha_3 a_1a_2 } \delta^{A_1 A_2} 
\end{array} \\
\begin{array}{c} \includegraphics[width=0.3\textwidth]{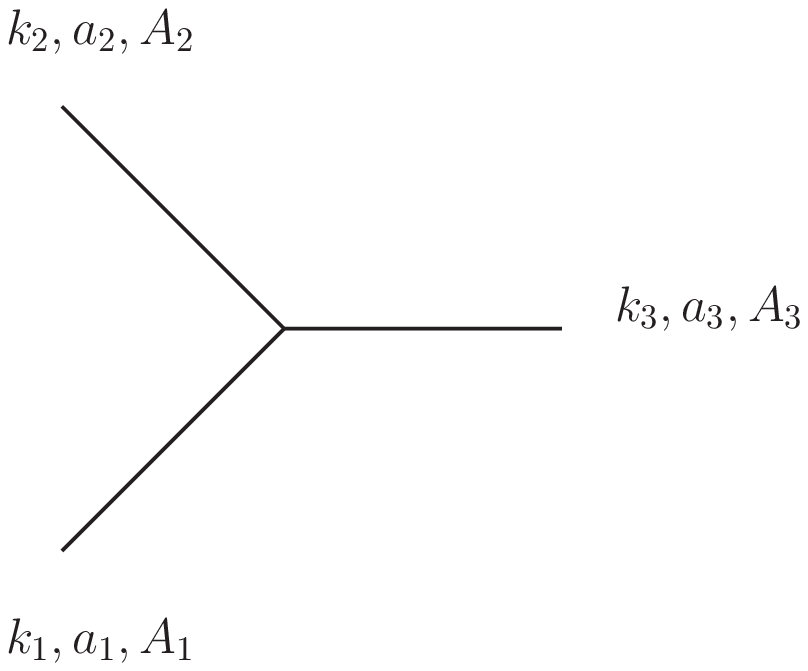} \end{array} &=&
\begin{array}{l}  i g \lambda f^{a_1a_2a_3} \hat f^{A_1A_2A_3}  
\end{array} \\
\begin{array}{c} \includegraphics[width=0.3\textwidth]{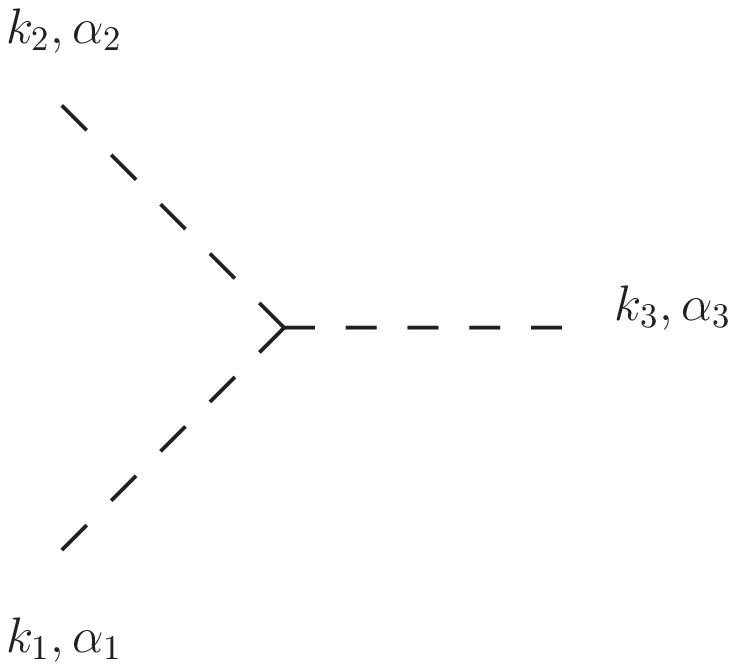} \end{array} &=&
\begin{array}{l}  i g  d^{\alpha_1 \alpha_2 \alpha_3}   
\end{array} 
\end{eqnarray}

\begin{eqnarray}
\begin{array}{c}\includegraphics[width=0.3\textwidth]{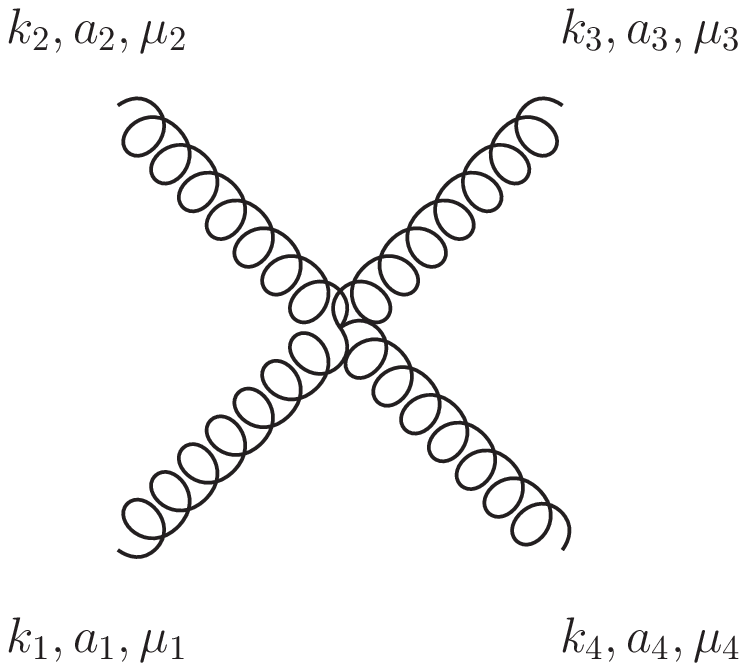} \end{array} &=& 
\begin{array}{l} \\ \\ \\ \\[10pt]
-i g^2 f^{a_1a_2b}f^{ba_3a_4} \Big\{ {1 \over 8 }(k_1-k_2)\cdot(k_3-k_4) \eta^{\mu_1 \mu_2} \eta^{\mu_3 \mu_4} \\
\hskip 0.5cm + {1 \over 4 }\big((k_1+k_2)^2+ (k_3+k_4)^2 - m^2 \big) \eta^{\mu_1 \mu_4} \eta^{\mu_2 \mu_3}  \\ 
\hskip 0.5cm - \Big({1 \over 2}k_4^{\mu_1} k_4^{\mu_4} + 2 k_4^{\mu_1}k_3^{\mu_4} +2 k_2^{\mu_1}k_4^{\mu_4} +k_4^{\mu_1}k_4^{\mu_4} \Big) \eta^{\mu_2 \mu_3} \\
\hskip 0.5cm + k_1^{\mu_1}(2k_1 + k_2 + k_3 )^{\mu_2}\eta^{\mu_3 \mu_4} + 2 k_2^{\mu_1} k_4^{\mu_3} \eta^{\mu_2 \mu_4} \Big\} \\ 
\hskip 0.5cm + \text{Perms}(1,2,3,4) \end{array} \nonumber\\ \\
\begin{array}{c}\includegraphics[width=0.3\textwidth]{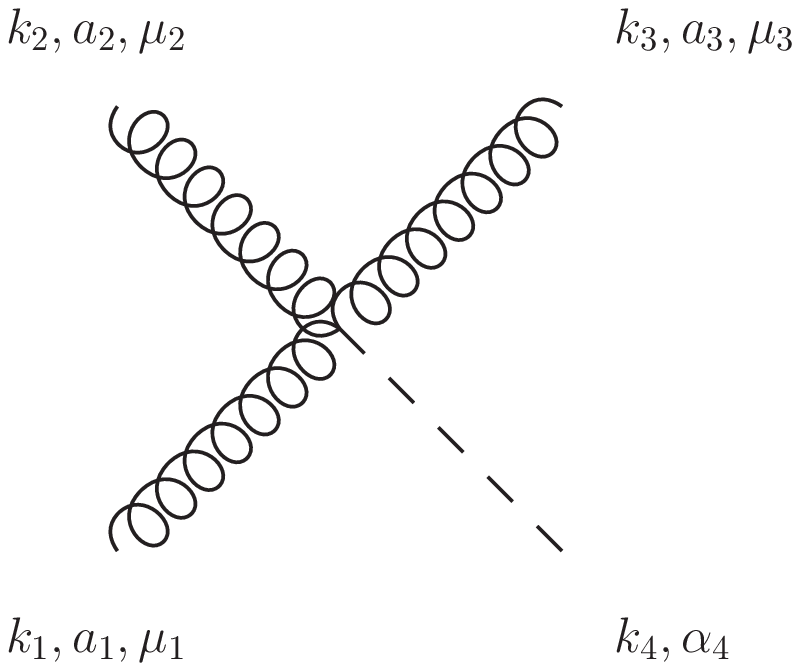} \end{array} &=&
\begin{array}{l} 2 g^2 f^{a_1a_2 b} C^{ \alpha_4 b a_3}  \eta^{\mu_1 \mu_3}  k_3^{\mu_2} + \text{Perms}(1,2,3) 
\end{array} \\
\begin{array}{c}\includegraphics[width=0.3\textwidth]{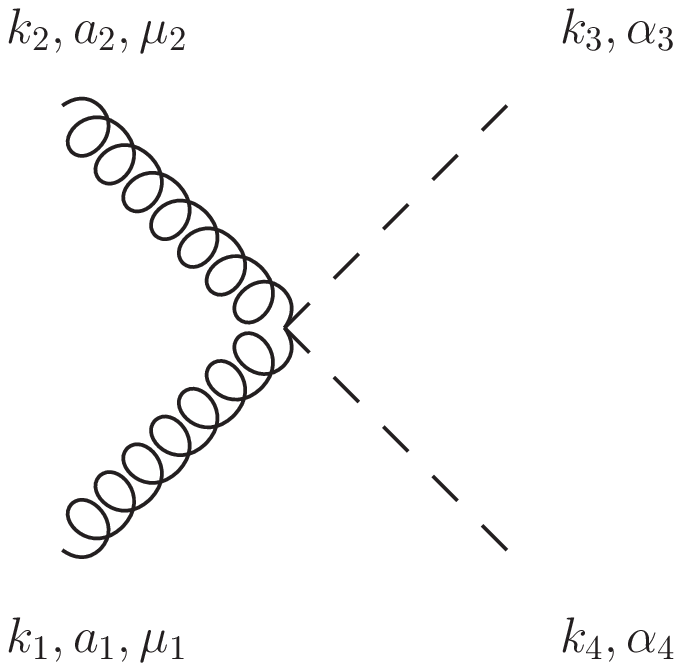} \end{array} &=&
\begin{array}{l} i g^2 (T_R^{a_1}T_R^{a_2}+T_R^{a_2}T_R^{a_1})^{\alpha_3 \alpha_4} + \text{Perms}(3,4) 
\end{array} \\
\begin{array}{c}\includegraphics[width=0.3\textwidth]{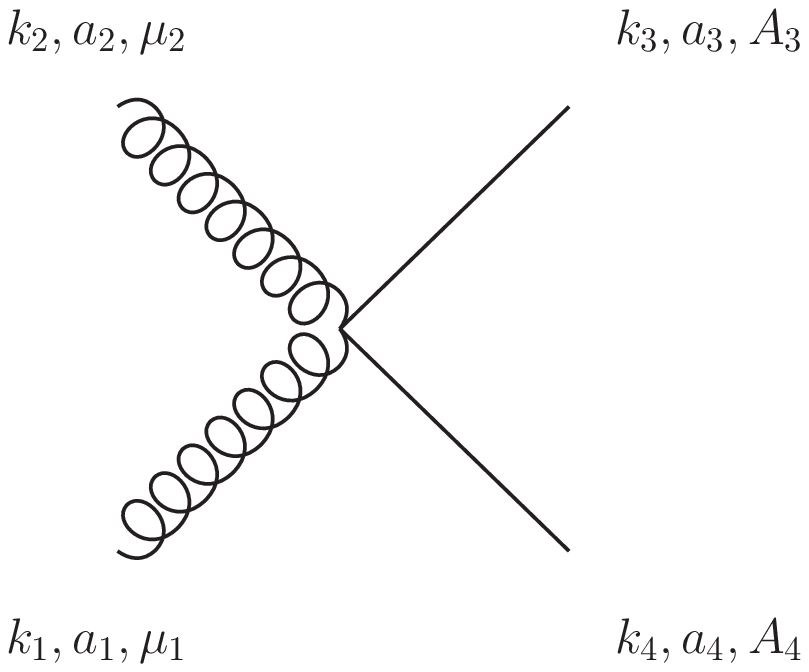} \end{array} &=&
\begin{array}{l}  i g^2 (f^{a_1a_4 b} f^{b a_2a_3} + f^{a_1a_3 b} f^{b a_2a_4} )\delta^{A_1 A_2} \eta^{\mu_1 \mu_2} 
\end{array} 
\end{eqnarray}

Expressions for five- and six-vertices can be found in ref.\ \cite{Johansson:2017srf}. 
Partial amplitudes are obtained introducing gauge-group generators (canonically) normalized as  
\begin{equation}
f^{a_1a_2a_3} = - {i } \text{Tr} ([T^{a_1}, T^{a_2}] T^{a_3}) \ ,
\end{equation}
where $f^{a_1a_2a_3}$ are real structure constants.

\subsection{Five-point examples of $(DF)^2+{\rm YM}+\phi^3$ amplitudes}
\label{appA1}

In this appendix, we display five-point examples of the kinematic factors $A_{(DF)^2+{\rm YM}+\phi^3}$ in heterotic-string
amplitudes (\ref{4.4}). Five scalars of the $(DF)^2+{\rm YM}+\phi^3$ theory give rise to
\begin{align}
A_{(DF)^2+{\rm YM}+\phi^3}&(1_s,2_s,3_s,4_s,5_s) =   i  \frac{ \lambda^3 \hat f^{A_1 A_2 B}\hat f^{BA_3 C}\hat f^{CA_4 A_5} }{4  s_{12} s_{45}} 
+ i  \frac{ \lambda \alpha' \delta^{A_1 A_3}\hat f^{A_2 A_4 A_5} }{ s_{45}  (1-2\alpha's_{13})  } 
  \notag \\
& + i  \frac{ \lambda \alpha'  \delta^{A_1 A_2}\hat f^{A_3 A_4 A_5}  }{ (1-2\alpha's_{12}) } \Big(  \frac{ s_{13}  }{ s_{12} s_{45} }{+}\frac{ s_{25} }{s_{12}s_{34}} {-}\frac{  1 }{s_{12} } \Big)+ {\rm cyc}(1,2,3,4,5) \ ,
\end{align}
and three scalars yield 
\begin{align}
&A_{(DF)^2+{\rm YM}+\phi^3}(1_s,2_g,3_g,4_s,5_s) =   i  \lambda \hat f^{A_1 A_4 A_5}   \bigg\{ 
\! \frac{  (e_2{\cdot }k_1) (e_3{\cdot }k_4)}{s_{12} s_{34}}
\! -  \! \frac{(e_2{\cdot }k_1)(e_3 {\cdot } k_{12})}{s_{12} s_{45}} 
\! - \! \frac{(e_2{\cdot} k_{34} ) (e_3{\cdot }k_4)}{s_{34} s_{15}}
 \notag \\
& \ \ 
- \frac{ (e_2{\cdot }k_1)(e_3{\cdot }k_2) -   (e_3{\cdot }k_1)(e_2{\cdot }k_3)+ s_{1 3} (e_2{\cdot }e_3)}{s_{23} s_{45}}
- \frac{ (e_2{\cdot }k_3)(e_3{\cdot }k_4)- (e_3{\cdot }k_2)(e_2{\cdot }k_4)+ s_{2 4} (e_2{\cdot }e_3)}{s_{23} s_{15}}
 \notag \\
& \ \ + \frac{ (e_2{\cdot }e_3)}{s_{23}}+ 2\alpha' \frac{   (e_2{\cdot }k_3) (e_3{\cdot }k_2) - s_{23} (e_2{\cdot }e_3)}{1 - 2\alpha' s_{23}} \Big( \frac{ s_{2 4} }{s_{23} s_{15}}
 + \frac{ s_{1 3}}{s_{23} s_{45}} - \frac{1}{s_{23}} \Big)
\bigg\} \ ,
\end{align}
where the $\alpha'$-dependent terms in the last line are separately gauge invariant. With two scalars and three gluons, the variety of tensor structures can be conveniently represented using the notation ${\cal E}_{ij} = (e_i\cdot e_j) + 2\ap \frac{ s_{ij} (e_i \cdot e_j ) - (e_i{\cdot }k_j)(e_j{\cdot }k_i) }{1-2\ap s_{ij}} =  \frac{ (e_i\cdot e_j)  -   2\ap (e_i{\cdot }k_j)(e_j{\cdot }k_i)}{1-2\ap s_{ij}}$ from \cite{Schlotterer:2016cxa}
\begin{align}
&A_{(DF)^2+{\rm YM}+\phi^3}(1_g,2_g,3_g,4_s,5_s) = 2  \delta^{A_4 A_5}  \bigg\{ 
\frac{(e_1 {\cdot} k_5) \big[(e_2 {\cdot} k_4) (e_3 {\cdot} k_2) - (e_2 {\cdot} k_3) (e_3 {\cdot} k_4) \big] }{s_{15} s_{23}}  \notag \\
&+ \frac{ (e_3 {\cdot} k_4) \big[ (e_2 {\cdot} k_1) (e_1 {\cdot} k_5) - (e_2 {\cdot} k_5) (e_1 {\cdot} k_2) \big] }{s_{12} s_{34}}
- \frac{ (e_2{\cdot} k_{34}) (e_1 {\cdot} k_5) (e_3 {\cdot} k_4)}{s_{15} s_{34}} \notag \\
&+ \frac{ (e_3{\cdot}k_{12}) \big[(e_1 {\cdot} k_4) (e_2 {\cdot} k_1) - (e_1 {\cdot} k_2) (e_2 {\cdot} k_4)\big]+ (e_3 {\cdot} k_4) \big[(e_1 {\cdot} k_2) (e_2 {\cdot} k_3) - (e_1 {\cdot} k_3) (e_2 {\cdot} k_1)\big] }{s_{12} s_{45}} \notag \\
&- \frac{(e_1{\cdot}k_{23}) \big[(e_3 {\cdot} k_5) (e_2 {\cdot} k_3) - (e_3 {\cdot} k_2) (e_2 {\cdot} k_5)\big] 
+ (e_1 {\cdot} k_5) \big[(e_3 {\cdot} k_2) (e_2 {\cdot} k_1) - (e_3 {\cdot} k_1) (e_2 {\cdot} k_3)\big] }{s_{23} s_{45}}
\notag \\
&+{\cal E}_{12} \Big[(e_3 {\cdot} k_4) \Big( \frac{s_{25}}{s_{12} s_{34}} - \frac{ s_{23}}{s_{12} s_{45}} - \frac{ 1}{s_{45}} \Big) 
+\frac{ (e_3 {\cdot} k_1) s_{24}}{s_{12} s_{45}} - \frac{ (e_3 {\cdot} k_2) s_{14}}{s_{45}} \Big( \frac{1}{s_{12}} + \frac{1}{s_{23}} \Big)  \Big]  \\
&+ {\cal E}_{23} \Big[ (e_1 {\cdot} k_5) \Big( \frac{ s_{12}}{s_{23} s_{45}} - \frac{ s_{24}}{s_{15} s_{23}} + \frac{1}{s_{45}} \Big) 
 - \frac{(e_1 {\cdot} k_3) s_{25}}{s_{23} s_{45}} + \frac{ (e_1 {\cdot} k_2) s_{35}}{s_{45}} \Big(\frac{1}{s_{12}} +\frac{ 1}{s_{23}}\Big) \Big] \notag \\
 & + {\cal E}_{13} \Big[ \frac{ e_2 {\cdot} k_4}{s_{45}} + \frac{(e_2 {\cdot} k_1) (s_{34} + s_{45})}{s_{12} s_{45}} +\frac{ (e_2 {\cdot} k_3) s_{14}}{s_{23} s_{45}} \Big]
 \notag \\
&+ \frac{2\ap}{1-2\ap s_{45}} \Big( \frac{ s_{14} }{s_{23}s_{45} } + \frac{ s_{35}}{s_{12} s_{45}} - \frac{1}{s_{45}} \Big) \Big( (e_2{\cdot} k_1)(e_1{\cdot} k_3)(e_3{\cdot} k_2) - (e_2{\cdot} k_3)(e_3{\cdot} k_1)(e_1{\cdot} k_2)  \notag \\
& \ \ 
+ {\cal E}_{12} \big[ s_{23}(e_3 {\cdot} k_1){-}s_{13}(e_3 {\cdot} k_2)   \big]
+ {\cal E}_{23} \big[ s_{13}(e_1 {\cdot} k_2){-}s_{12}(e_1 {\cdot} k_3)   \big]
+ {\cal E}_{13} \big[ s_{12}(e_2 {\cdot} k_3){-}s_{23}(e_2 {\cdot} k_1)   \big]
\Big)
\bigg\} \, . \notag
\end{align}
Finally, the five-gluon case $A_{(DF)^2+{\rm YM}+\phi^3}(1_g,2_g,3_g,4_g,5_g) =-(\ap)^{-1}B(1,2,3,4,5) $  is available from the arXiv submission of ref.~\cite{Huang:2016tag}.

\subsection{BCJ relations of $(DF)^2+{\rm YM}+\phi^3$ amplitudes}
\label{appA2}

We shall now derive the BCJ relations (\ref{4.6}) of the kinematic factors $A_{(DF)^2+{\rm YM}+\phi^3}$ in (\ref{4.4}). The idea is to study the non-supersymmetric chiral half of the underlying heterotic-string correlators involving the currents $J^A(z)$ and $\partial X^\mu(z)$ in an open-string setup. Such disk correlators can be realized by augmenting the open bosonic string by vertex operators $V^A(z) = J^A(z) e^{ik\cdot X(z)}$ of massless scalars which carry an additional adjoint degree of freedom $A$ besides their Chan--Paton generators $T^a$. Such vertex operators may be obtained from compactifications of the open bosonic string in the geometries known from the non-supersymmetric chiral half of the heterotic string and yield tree amplitudes
\beq
 {\cal A}^{\rm comp}_{\rm bos}(1,\pi(2,3,\ldots,n{-}2),n{-}1,n) = \! \sum_{\rho \in S_{n-3}} \! \! F_\pi{}^\rho A_{(DF)^2+{\rm YM}+\phi^3}(1,\rho(2,3,\ldots,n{-}2),n{-}1,n) \, ,
 \label{A.X1double}
\eeq
cf.\ (\ref{A.X1}). Descending from a worldsheet of disk topology, the color-ordered amplitudes (\ref{A.X1double}) involving gluons and scalars satisfy the monodromy relations \cite{BjerrumBohr:2009rd, Stieberger:2009hq}. Following the ideas of ref.~\cite{Broedel:2012rc, Huang:2016tag}, one may restrict the $\ap$-expansion of these monodromy relations to their contributions of lowest transcendentality. Their imaginary parts then degenerate to BCJ relations with $F_\pi{}^\rho \rightarrow \delta_\pi^\rho$ and yield the desired property (\ref{4.6}) of $A_{(DF)^2+{\rm YM}+\phi^3}$. Upon inserting the Z-amplitude expansion (\ref{2.15}) of the integrals $F_\pi{}^{\rho}$, (\ref{A.X1}) gives rise to the double-copy construction of the compactified open bosonic string noted in table \ref{overview}.

\section{Derivation of explicit amplitude relations for conformal supergravity}
\label{appB}

\subsection{CHY derivation of the one- and two-graviton relations}
\label{appB1}

In this section, we will derive the amplitude relations (\ref{4.2new}) and (\ref{4.3new}) for conformal
supergravity coupled to gauge bosons from their representation in the CHY formalism.
To begin with, recall the CHY integrand for an $n$-point color-ordered amplitude of the gauge multiplet,
\begin{equation}
\mathcal{I}_n^{\mathrm{SYM}}(1,2,\ldots,n)=\mathcal{C}(1,2,\ldots,n)\,\mathrm{Pf}^\prime\,\Psi_n(\{k,e,z\})\,,
\label{chyym}
\end{equation}
where $\mathcal{C}(1,2,\ldots,n) \equiv (z_{12}z_{23}\ldots z_{n1})^{-1}$ is the Parke--Taylor factor. The reduced Pfaffian Pf${}^\prime\,\Psi_n(\{k,e,z\})$ of the usual CHY matrix depending on the momenta $k_j$, the polarizations $e_j$ and the complex coordinates $z_j$ on the Riemann sphere as well as the integration prescription can be found in ref.~\cite{Cachazo:2013hca}. The Pfaffian may be supersymmetrized via pure-spinor methods \cite{Berkovits:2013xba}, and the result coincides with the $n$-point correlators of the open pure-spinor superstring \cite{Mafra:2011kj, Mafra:2011nv} after stripping off the Koba--Nielsen factor (see \cite{Gomez:2013wza} for the detailed argument). The expression (\ref{chyym}) can be generalized to the single-trace sector of an $r$-graviton, $n$-gluon EYM amplitude \cite{Cachazo:2014nsa},
\begin{equation}
\mathcal{I}_{n+r}^{\mathrm{EYM}}(1,2,\ldots,n;p_1,\ldots,p_r)=\mathcal{C}(1,2,\ldots,n)\,\mathrm{Pf}\,\Psi_r(\{p,e_p,z\})\,\mathrm{Pf}^\prime\,\Psi_{n+r}(\{k,p,e_k,e_p,z\})\,,
\label{EYM.CHY}
\end{equation}
with graviton momenta $p$ and polarizations $e_p$. These expressions were used in ref.~\cite{Nandan:2016pya} 
to prove relations equivalent to the $\ap \to 0$ limit of (\ref{4.2}) and (\ref{4.3}). Note that the $r$-particle Pfaffian
$\mathrm{Pf}\,\Psi_r(\{p,e_p,z\})$ in (\ref{EYM.CHY}) cannot be supersymmetrized.

For the theory of conformal supergravity coupled to gauge bosons, there is an expression analogous to (\ref{EYM.CHY}) which, as is implicit in  \cite{Azevedo:2017lkz}, can be obtained from the heterotic ambitwistor string and is  given by  
\begin{equation}
\mathcal{I}_{n+r}^{R^2\mathrm{YM}}(1,2,\ldots,n;p_1,\ldots,p_r)=\mathcal{C}(1,2,\ldots,n) \left. W_{\underbrace{11\cdots1}_r}(\{p,e_p,z\})\right.\mathrm{Pf}^\prime\,\Psi_{n+r}(\{k,p,e_k,e_p,z\})\,,
\label{R2YM.CHY}
\end{equation}
where $W_{\{I\}}$ denotes a product of the Lam--Yao cycles. The latter have been introduced in ref.~\cite{Lam:2016tlk} and were later used in ref.~\cite{He:2016iqi} in the study of the CHY formulation of amplitudes coming from theories involving higher-dimensional operators, such as $F^3$ for gauge theory and $R^2$ for gravity. In the particular case appearing in (\ref{R2YM.CHY}), it can be written in terms of the diagonal elements of the $C$-block of $\Psi_{n+r}$ as
\begin{equation}
W_{\underbrace{11\cdots1}_r}(\{p,e_p,z\}) = \prod_{i=1}^r C_{p_ip_i} \equiv \prod_{i=1}^r \bigg(\sum_{j=1}^n  \frac{e_{p_i}\cdot k_j}{z_j-z_{p_i}} + \sum_{\substack{j=1\\(j\neq i)}}^{r} \frac{e_{p_i}\cdot p_j}{z_{p_j}-z_{p_i}}\bigg) \, .
\label{seehere}
\end{equation}
We are now ready to prove that (\ref{4.2}) literally translates into (\ref{4.2new}) for conformal gravity. In fact, since Pf$\,\Psi_{r=1} = W_1 = C_{pp}$, the proof is identical to the one given in ref.~\cite{Nandan:2016pya}, and amounts to showing that, in this case,
\begin{equation}
C_{pp} = \sum_{j=1}^{n-1} (e_p \cdot x_j) \frac{z_{j,j+1}}{z_{j,p}z_{p,j+1}}\,,
\end{equation}
with $x_j=k_1{+}k_2{+}\ldots{+}k_j$, see the reference for details. The similarity of the results in both theories is of course not surprising since (\ref{4.2}) has no explicit dependence on $\ap$.

Moving on to the case with two gravitons,  writing $p_1 \equiv p$ and $p_2 \equiv q$, one has
\begin{eqnarray}
W_{11} & = & C_{pp}C_{qq} \nonumber\\
 & = &  C^\prime_{pp}C^\prime_{qq} +  C^\prime_{pp}(e_q\cdot p)\frac{z_{pn}}{z_{pq}z_{qn}} +  C^\prime_{qq}(e_p\cdot q)\frac{z_{qn}}{z_{qp}z_{pn}} - \frac{(e_p\cdot q)(e_q\cdot p)}{z_{pq}^2}\,,   \label{W11}
\end{eqnarray}
where we have defined $C^\prime_{pp} \equiv \sum_{i=1}^{n-1} (e_p\cdot x_i) z_{i,i+1}/(z_{i,p}z_{p,i+1})$ and written
\begin{equation}
C_{pp} = C^\prime_{pp} + (e_p\cdot q)\frac{z_{qn}}{z_{qp}z_{pn}}\,.
\end{equation}
Analogous expressions can be written for $C_{qq}$ by swapping $p$ and $q$. Now, comparing with the expression for the Pfaffian in (\ref{EYM.CHY}),
\begin{equation}
\mathrm{Pf}\,\Psi_{r=2} = C^\prime_{pp}C^\prime_{qq} +  C^\prime_{pp}(e_q\cdot p)\frac{z_{pn}}{z_{pq}z_{qn}} +  C^\prime_{qq}(e_p\cdot q)\frac{z_{qn}}{z_{qp}z_{pn}} - \frac{s_{pq}(e_p\cdot e_q)}{z_{pq}^2}\,,
\label{Pf2}
\end{equation}
we see that again we can benefit from the work done in the EYM case. Indeed, one immediately sees that the manipulations involving the tensor structures $(e_p\cdot x_i)(e_q\cdot x_j)$ and $(e_p\cdot x_i)(e_q\cdot p)$ will be exactly the same in the conformal gravity theory, which implies that both the first and the third lines of (\ref{4.3}) apply in this case. Once more, this could be expected from the absence of explicit $\ap$-dependence in those terms.

The last term in (\ref{W11}) has a new tensor structure w.r.t. the Pfaffian. However, it is very similar in form to the last term in (\ref{Pf2}). In fact, the former can be obtained from the latter by replacing $(e_p\cdot e_q)$ with $s_{pq}^{-1}(e_p\cdot q)(e_q\cdot p)$. This in turn ultimately implies that one can perform such a substitution directly in the expression relating the (single-trace) scattering amplitudes of two gravitons and $n$ gluons in EYM to that of $n{+}2$ gluons, and thus obtain the analogous expression in conformal supergravity. Doing so in the expression given in ref.~\cite{Nandan:2016pya}, we get an additive contribution of
\begin{equation}
-\frac{(e_p\cdot q)(e_q\cdot p)}{2 s_{pq}} \sum_{l=1}^{n-1} (p \cdot k_l)  \sum_{1=i\leq j}^l 
A_{\text{SYM}}( 1,2,\ldots, i{-}1,q,i, \ldots,j{-}1,p,j,\dots ,n) + (  p \leftrightarrow q) \,, 
\end{equation}
which has precisely the same form as the $\ap \to \infty$ limit of the second line of (\ref{4.3}). This reproduces the second line of (\ref{4.3new}) and completes the proof that  the $\ap \rightarrow \infty$ limit of the heterotic-string relations (\ref{4.2}) and (\ref{4.3})  applies to conformal supergravity coupled to gauge bosons.

\subsection{The three-graviton relation}
\label{appB2}

In order to conveniently represent amplitudes $A_{{\rm CSG}+{\rm SYM}}(1,2, \ldots,n; p,q,r)$
involving three gravitons and $n$ gluons in terms of gauge-theory amplitudes, we recursively define the shuffle product of ordered sets $P\equiv \{p_1,p_2,\ldots,p_m\}$ and $Q\equiv \{q_1,q_2,\ldots,q_n\}$ of cardinality $m$ and $n$, respectively:
\beq
P \shuffle \emptyset =  \emptyset \shuffle P = P \ , \ \ \ \ 
P \shuffle Q= \{p_1(p_2,\ldots,p_m \shuffle Q) \} + \{q_1(q_2,\ldots,q_n) \shuffle P\} \, .
\label{shuf}
\eeq
Then, the notation $\sum_{\sigma\in P \shuffle Q}$ instructs to sum over all permutations $\sigma$ of $P \cup Q$ which preserve the individual orderings of $P$ and $Q$. This applies to the $\ap \rightarrow \infty$ limit of the three-graviton relation in ref.~\cite{Schlotterer:2016cxa}
\begin{align}
 &A_{{\rm CSG}+{\rm SYM}}(1,2, \ldots,n; p,q,r) = \frac{ (e_p\cdot q)(e_q\cdot p)(e_r \cdot q) }{s_{pq}}\sum_{j=1}^{n-1} s_{jp}\sum_{\sigma \in \{r,q,p\} \atop{\shuffle \{ 1,2,\ldots, j-1\}} } A_{\rm SYM}(\sigma,j,j{+}1,\ldots,n) \notag \\
&-\, \frac{  (e_p\cdot q)(e_q\cdot p)}{2 s_{pq}} \sum_{j=1}^{n-1}  (e_r \cdot x_j)  \Big\{
\sum_{i=1}^j s_{ip} \! \! \! \! \! \sum_{\sigma \in \{q,p\} \atop{\shuffle \{ 1,2,\ldots, i-1\}} } \! \! \! \! \! A_{\rm SYM} (\sigma,i,i{+}1,\ldots,j,r,j{+}1,\ldots,n)
\notag  \\
& \ \ \ \ \ \  \ \ \ \ \ \  + s_{pr} \sum_{\sigma \in \{q,p\} \atop{\shuffle \{ 1,2,\ldots, j\}} } A_{\rm SYM}(\sigma,r,j{+}1,\ldots,n)
+ \sum_{i=j+1}^{n-1} s_{ip} \! \! \! \! \!\! \! \! \! \! \sum_{\sigma \in \{q,p\} \atop{\shuffle \{ 1,\ldots,j,r,j{+}1,\ldots i-1\}} }\! \! \! \! \! \! \! \! \! \! A_{\rm SYM} (\sigma,i,i{+}1,\ldots,n) \Big\} \notag \\
& -  ( e_r \cdot p) \Big\{
\sum_{1=j\leq i}^{n-1}  (e_p \cdot x_i)  (e_q \cdot x_j) \! \! \! \! \! \sum_{\sigma \in \{r\} \atop{\shuffle \{ 1,\ldots,j,q,j{+}1,\ldots, i\}} } \! \! \! \! \! A_{\rm SYM} (\sigma,p,i{+}1,\ldots,n) 
  \label{threegr} \\
& \ \ \ \ \ \  \ \ \ \ \ \ + \sum_{1=i\leq j}^{n-1}  (e_p \cdot x_i)  (e_q \cdot x_j)  \sum_{\sigma \in \{r\} \atop{\shuffle \{ 1,2,\ldots, i\}} } A_{\rm SYM} (\sigma,p,i{+}1,\ldots,j ,q,j{+}1,\ldots,n) 
\Big\} \notag \\
&+\sum_{1=i\leq j\leq k}^{n-1}  (e_p \cdot x_i)  (e_q \cdot x_j)(e_r \cdot x_k)    A_{\rm SYM} (1,2,\ldots,i,p,i{+}1,\ldots,j,q,j{+}1,\ldots,k,r,k{+}1,\ldots,n)\notag \\
& + (e_p \cdot (q+r)) ( e_q \cdot r) \sum_{j=1}^{n-1}  (e_r \cdot x_j)  \sum_{\sigma \in \{p,q\} \atop{\shuffle \{ 1,2,\ldots, j\}} } A_{\rm SYM} (\sigma,r,j{+}1,\ldots,n)  \notag \\
&+ \frac{1}{3}  \,  {\cal F}_{pqr} \,
 \sum_{j=1}^{n-1} s_{jr} \! \! \!  \sum_{ \sigma \in \{ p,q,r\} \atop { \shuffle \{ 1,2,\ldots,j-1 \} } } \! \! \!  A_{\rm SYM} (\sigma,j,j{+}1,\ldots,n)
  +{\rm perm}(p,q,r)\notag
\end{align}
with non-local kinematic factors
\begin{align}
&{\cal F}_{pqr} = \frac{1}{s_{pqr}} \bigg\{ (e_p\cdot q)(e_q\cdot r)(e_r\cdot p) - (e_p\cdot r)(e_r\cdot q)(e_q\cdot p) \label{calFF} \\
&+ \Big[ \frac{ (e_p\cdot q)(e_q\cdot p) }{s_{pq}} 
\big( s_{pr}(e_r\cdot q) - s_{qr}(e_r\cdot q) \big) + {\rm cyc}(p,q,r)
\Big] \, \bigg\}  \, . \notag
\end{align}
The methods of appendix \ref{appB1} allow for a direct derivation of (\ref{threegr}) from the CHY formalism.

\subsection{The double-trace relation}
\label{appB3}

The double-trace sector of the $n$-gluon amplitude $A_{{\rm CSG}+{\rm SYM}}(\{ 1,2,\ldots,r \, | \,r{+}1,\ldots,n \})$ can also be related to (single-trace) gauge-theory amplitudes. The CHY integrand piece associated with the double-trace sector of ${\rm Tr}(T^{a_1} T^{a_2} \ldots T^{a_r}) {\rm Tr}(T^{a_{r+1}} \ldots T^{a_n})$ is  (up to a sign) just a product of the corresponding Parke--Taylor factors:
\begin{equation}
\mathcal{I}_{r,n-r}^{R^2\mathrm{YM}}(\{ 1,2,\ldots,r \, | \,r{+}1,\ldots,n \})=-\,\mathcal{C}(1,2,\ldots,r)\, \mathcal{C}(r{+}1,\ldots,n)\,\mathrm{Pf}^\prime\,\Psi_{n}(\{k,e,z\})\,.
\label{R2YM.CHY.doubletrace}
\end{equation}
Again, this comes naturally from ambitwistor-string correlators of gluon vertex operators, which involve the same current-algebra OPEs as in ordinary heterotic string theory \cite{Frenkel:2010ys, Dolan:2007eh}.

The EYM version of (\ref{R2YM.CHY.doubletrace}) is given by \cite{Cachazo:2014nsa}
\begin{equation}
\mathcal{I}_{r,n-r}^{\mathrm{EYM}}(\{ 1,2,\ldots,r \, | \,r{+}1,\ldots,n \})= s_{12\cdots r} \, \mathcal{C}(1,2,\ldots,r)  \,\mathcal{C}(r{+}1,\ldots,n)\,\mathrm{Pf}^\prime\,\Psi_{n}(\{k,e,z\})\,,
\label{EYM.CHY.doubletrace}
\end{equation}
where $s_{12\cdots r} \equiv \frac{1}{2}(k_1 + k_2 + \cdots + k_r)^2$. Comparing (\ref{R2YM.CHY.doubletrace}) with (\ref{EYM.CHY.doubletrace}), it is easy to see that one can recycle the EYM double-trace amplitude relations derived in ref.~\cite{Nandan:2016pya} to obtain the analogous relations for conformal gravity. Indeed, it suffices to multiply the right-hand side of the former by $-1/s_{12\cdots r}$. The desired expression is thus
\begin{align}
&A_{{\rm CSG}+{\rm SYM}}(\{ 1,2,\ldots,r \, | \,r{+}1,\ldots,n \}) =     \label{CSGdoubletracefinal}
 \\
& \ \ \ \ \  \frac{1}{s_{12\cdots r} }\, \sum_{j=1}^{r-1} \sum_{\ell=r+2}^{n} (-1)^{n-\ell} s_{j\ell}  \sum_{\tau \in \{ r+2,\ldots, \ell-1 \} \atop { \shuffle \{n,n-1,\ldots,\ell+1 \} } } \sum_{\sigma \in \{1,2,\ldots, j-1\} \atop{ \shuffle \{r+1,\tau,\ell \} }} \! \! A_{\rm SYM}(\sigma,j,j{+}1,\ldots,r) \,,
\notag
\end{align}
which, as expected, has the same form as the $\ap \rightarrow \infty$ limit of the corresponding (ordinary) heterotic-string relations for double-trace amplitudes \cite{Schlotterer:2016cxa}   
\begin{align}
&{\cal A}_{\rm het}^{(2)}(\{ 1,2,\ldots,r \, | \,r{+}1,\ldots,n \}) =   -  \, \frac{2\alpha'}{1-2\alpha' s_{12\ldots r} }  
  \label{CSGdoubletracefinal2}
 \\
& \ \ \ \ \  \times \sum_{j=1}^{r-1} \sum_{\ell=r+2}^{n} (-1)^{n-\ell} s_{j\ell}  \sum_{\tau \in \{ r+2,\ldots, \ell-1 \} \atop { \shuffle \{n,n-1,\ldots,\ell+1 \} } } \sum_{\sigma \in \{1,2,\ldots, j-1\} \atop{ \shuffle \{r+1,\tau,\ell \} }} \! \! {\cal A}_{\rm het}(\sigma,j,j{+}1,\ldots,r) \,.
\notag
\end{align}
Note that the ordered set $ \{r{+}1,\tau,\ell \}$ in summation range for $\sigma$ in (\ref{CSGdoubletracefinal}) and (\ref{CSGdoubletracefinal2}) is understood as inserting the permutation $\tau$ from the previous one in between legs $r{+}1$ and $\ell$. 

\subsection{The CHY integrand for ${\rm CG}+(DF)^2$}
\label{appB7}

The amplitude relations of EYM can be derived as a consequence of the CHY-half-integrand $\mathcal{C}(1,2,\ldots,n)\,\mathrm{Pf}\,\Psi_r(\{p,e_p,z\})$ and its multi-trace generalizations \cite{Cachazo:2014nsa, Cachazo:2014xea}. Hence, the amplitude relations of EYM are unchanged when  the second half-integrand $\mathrm{Pf}^\prime\,\Psi_{n+r}(\{k,p,e_k,e_p,z\})$ is replaced by a different ${\rm SL}(2,\mathbb{C})$ covariant object. Given that EYM and the ${\rm CG}+(DF)^2$ theory share a factor of $A_{{\rm YM} + \phi^3}$ in the KLT-form of their tree amplitudes, they must also share the corresponding CHY-half-integrand $\mathcal{C}(1,2,\ldots,n)\,\mathrm{Pf}\,\Psi_r(\{p,e_p,z\})$ and its multi-trace generalizations. This proves our claim in subsection \ref{sec44} on the matching of EYM and ${\rm CG}+(DF)^2$ amplitude relations. From the $(DF)^2$ constituent in its double-copy structure, the CHY integrand of the single-trace sector of the ${\rm CG}+(DF)^2$ theory is given by
\begin{equation}
\mathcal{I}_{n+r}^{R^2(DF)^2}(1,2,\ldots,n;p_1,\ldots,p_r)=\mathcal{C}(1,2,\ldots,n)\,\mathrm{Pf}\,\Psi_r(\{p,e_p,z\}) \left. W_{\underbrace{11\cdots1}_{n+r}}(\{k,p,e_k,e_p,z\})\right. 
\label{R2DF2.CHY}
\end{equation}
with the $W_{11\cdots1}$-factor defined in (\ref{seehere}).

\bibliographystyle{JHEP}
\bibliography{cites}{}

\end{document}